\documentclass{article}

\usepackage{arxiv}

\usepackage[utf8]{inputenc} 
\usepackage[T1]{fontenc}    
\usepackage{hyperref}       
\usepackage{url}            
\usepackage{booktabs}       
\usepackage{amsmath}        
\usepackage{amssymb}        
\usepackage{nicefrac}       
\usepackage{microtype}      
\usepackage{graphicx}       
\usepackage{natbib}         
\usepackage{doi}            
\usepackage{xcolor}         
\usepackage{CJK}            
\usepackage{enumerate}      
\usepackage{bm}             
\usepackage{indentfirst} 

\hypersetup{
  colorlinks=true,
  linkcolor=red,
  citecolor=orange,
  urlcolor=blue,
}

\graphicspath{{../figure}}

\usepackage{multicol} 

\usepackage{framed}
\usepackage{multirow}

\def\dt{\mbox{${\Delta t}$}}
\def\dx{\mbox{${\Delta x}$}}

\makeatletter
\@addtoreset{equation}{section}
\makeatother

\title{Energy-conserving finite difference scheme based on velocity interpolation 
applicable to unsteady flows using collocated grids
}

\author{\href{https://orcid.org/0000-0002-4875-8174}{\includegraphics[scale=0.06]{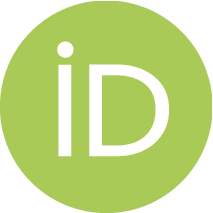}\hspace{1mm}Hideki Yanaoka}
\thanks{Email address for correspondence: yanaoka@iwate-u.ac.jp} \\
	Department of Systems Innovation Engineering, \\
    Faculty of Science and Engineering, Iwate University, \\
    4-3-5 Ueda, Morioka, Iwate 020-8551, Japan \\
}

\renewcommand{\shorttitle}{Energy-conserving finite difference scheme based on velocity interpolation 
applicable to unsteady flows}

\makeatletter
\hypersetup{
pdfusetitle,
bookmarksnumbered,
pdftitle={\@title},
pdfsubject={},
pdfauthor={Hideki Yanaoka},
pdfkeywords={Collocated grid, Kinetic energy conservation, Interpolation, Simultaneous relaxation, Incompressible flow, Finite difference method}
}
\makeatother

\begin{document}

\maketitle

\begin{abstract}
The collocation method uses the Rhie--Chow scheme to find the cell interface velocity 
by pressure-weighted interpolation. 
The accuracy of this interpolation method in unsteady flows has not been fully clarified. 
This study constructs a finite difference scheme for incompressible fluids 
using a collocated grid in a general curvilinear coordinate system. 
The velocity at the cell interface is determined by weighted interpolation 
based on the pressure difference to prevent pressure oscillations. 
The Poisson equation for the pressure correction value is solved 
with the cross-derivative term omitted to improve calculation efficiency. 
In addition, simultaneous relaxation of velocity and pressure is applied 
to improve convergence. 
Even without the cross-derivative term, calculations can be stably performed, 
and convergent solutions are obtained. 
In unsteady inviscid flow, the conservation of kinetic energy is excellent 
even in a non-orthogonal grid, 
and the calculation result has second-order accuracy to time. 
In viscous analysis at a high Reynolds number, 
the error decreases compared with that of the Rhie--Chow interpolation method. 
The present numerical scheme improves calculation accuracy in unsteady flows. 
The possibility of applying this computational method to high Reynolds number flows is demonstrated 
through several analyses.
\end{abstract}


\section{Introduction}

Numerical methods for incompressible flows usually use staggered grids 
to eliminate spurious errors for pressure. 
In this case, the calculation code becomes complicated 
because the definition points for each velocity component and pressure are different. 
For general curvilinear coordinate systems, 
it is hard to satisfy the conservation law discretely using staggered grids. 
In addition, it is not easy to set boundary conditions.

On the other hand, a method that does not use staggered grids has been proposed. 
\citet{Rhie&Chow_1983} eliminates pressure spurious errors 
by interpolating velocities with pressure gradient weights at cell interfaces. 
This method is called pressure interpolation or momentum interpolation. 
Later, \citet{Peric_et_al_1988} and \citet{Majumdar_1988} improved the interpolation method. 
\citet{Zang_et_al_1994} have extended the fractional-step method of \citet{Kim&Moin_1985} 
to general curvilinear coordinate systems using collocated grids. 
They adopted a method of interpolating the velocity by shifting upstream at the cell interface. 
The pressure interpolation of \citet{Rhie&Chow_1983} is known to have shortcomings. 
Improvement is necessary to obtain a convergent solution that does not depend on a time step \citep{Choi_1999}, 
and when the time step is small, pressure spurious errors occur \citep{Yu_et_al_2002}. 
\citet{Bartholomew_et_al_2018} proposed a unified and consistent formulation of Rhie--Chow's momentum interpolation 
and analyzed incompressible and low Mach number flows. 
\citet{Lee_et_al_2019} reported that the error of the continuity equation 
defined by the velocity at the cell center becomes second-order accuracy to time 
when interpolation by pressure difference is used. 
The numerical method using a collocated grid affects the conservation of kinetic energy 
\citep{Morinishi_1998, Morinishi_1999} 
even without using Rhie--Chow pressure interpolation. 
On the other hand, no studies refer to energy conservation when using pressure interpolation. 
The report of \citet{Lee_et_al_2019} does not concretely show the energy conservation property. 
To the best of the author's knowledge, 
the effect of interpolating the velocity at the cell interface on the time accuracy 
has not been investigated in the analysis of unsteady flows.

Numerical methods such as SIMPLE (Semi-Implicit Method for Pressure-Liked Equation) 
\citep{Patankar&Spalding_1972, Van_Doormaal&Raithby_1984} 
and MAC (Marker and Cell) \citep{Harlow&Welch_1965, Amsden&Harlow_1970, Hirt_et_al_1975} are used 
for analyzing incompressible flows. 
When the SIMPLE and MAC methods are expanded to general curvilinear coordinate systems, 
cross-derivative terms always appear in the fundamental equations. 
When the Poisson equation for pressure or pressure correction value is discretized 
by the central difference scheme with second-order accuracy, 
the coefficients of the Poisson equation with cross terms are 19 components. 
When analyzing a complicated three-dimensional flow field, 
the increase in memory usage and computation time results in inefficient computation. 
Therefore, it is necessary to construct a method that can perform stable and highly accurate calculations 
even if the cross terms are omitted. 
In particular, when solving the Poisson equation for pressure, 
the treatment method for the cross terms leads to computational instability. 
In numerical methods of the SIMPLE family, 
convergence and computational accuracy in the analysis 
without the cross terms have been investigated. 
\citet{Peric_1990} analyzed a two-dimensional cavity flow using a non-orthogonal grid system 
and found that omitting the cross term of the Poisson equation for the pressure correction value 
leads to slow convergence when using a highly strained grid. 
It is also revealed that in a distorted non-orthogonal grid system, 
the range of under-relaxation coefficient over which the calculation converges is very narrow. 
\citet{Wu_et_al_1995} reported that the omission of the cross term 
in the Poisson equation leads to significant differences with existing values.

To analyze unsteady flows with high accuracy using collocated grids, 
we should investigate the effect of interpolating the velocity at the cell interface 
on the time accuracy. 
It is also necessary to consider an efficient method of solving pressure. 
From the above points of view, in this research, 
we construct a finite difference scheme for incompressible flows 
in a general curvilinear coordinate system using a collocated grid 
and investigate kinetic energy conservation properties 
and an efficient method of finding pressure. 
The outline of the present numerical method is as follows: 
The SMAC (Simplified Marker and Cell) method \citep{Amsden&Harlow_1970}, 
which is a numerical method using staggered grids, 
is extended to collocated grid systems. 
We improve the pressure interpolation of \citet{Rhie&Chow_1983} 
so that the time discretization is the second-order accuracy in unsteady fields. 
In addition, we introduce the idea of HSMAC (Highly Simplified MAC) \citep{Hirt_et_al_1975} 
so that convergence does not deteriorate even if the cross terms are omitted. 
This method simultaneously relaxes the Poisson equation for the pressure correction value 
and the modified equations for velocity and pressure. 
Using the computational method proposed in this study, 
we analyze several flow fields and verify that the time discretization accuracy can be improved 
in unsteady flow fields.

The remainder of this paper is organized as follows: 
Section \ref{fundamental_equation} presents the fundamental equations 
and mentions the transport equation for kinetic energy and the conservation property. 
In Section \ref{transformation}, 
the governing equations in the Cartesian coordinate system are transformed 
into the general curvilinear coordinate system 
so that flow fields with arbitrary-shaped boundaries can be analyzed. 
In Section \ref{interpolation}, 
we investigate the accuracy of interpolating cell interface velocities 
and construct an improved method of Rhie--Chow's interpolation method. 
Section \ref{numerical_method} extends the interpolation method used in this study 
to general curvilinear coordinate systems 
and proposes a simultaneous relaxation method to solve the governing equations. 
Section \ref{verification} analyzes several flow models using this numerical method 
and investigates the kinetic energy conservation properties and computational accuracy. 
Finally, Section \ref{summary} presents a summary of the results.

\begin{table}[t]
\begin{framed}
\noindent
\centering{\bf Nomenclature}
\begin{multicols}{2}
\begin{tabbing}[t]
\hspace{5mm} \= \kill
  $e_i$ \> unit vector in direction of gravity \\
  $g$ \> gravity acceleration, m/s$^2$ \\
  $g_{mn}$ \> metric tensor \\
  $J$ \> Jacobian \\
  $k$ \> thermal conductivity, W/(m K) \\
  $K$ \> kinetic energy, J/kg \\
  $Nu$ \> Nusselt number \\
  $p$ \> pressure, Pa \\
  $Pr$ \> Prandtl number \\
  $Ra$ \> Rayleigh number \\
  $t$ \> time, s \\
  $T$ \> temperature, K \\
  $u_i$ \> flow velocity $(u, v, w)$, m/s \\
  $U_i$ \> cell interface velocity, m/s \\
  $U_m$ \> contravariant velocity \\
  $x_i$ \> coordinate $(x, y, z)$, m \\
  \> \\
  {\it Greek symbols} \\
  $\alpha$ \> thermal diffusivity coefficient, m$^2$/s \\
  $\beta$ \> volume expansion coefficient, 1/K \\
  $\delta_1, \delta_2$ \> difference operator \\
  $\Delta t$ \> time increment, s \\
  $\Delta T$ \> temperature difference, K \\
  $\Delta x_i$ \> grid width, m \\
  $\lambda$ \> $\lambda = 1, 2$ \\
  $\mu$ \> viscosity coefficient, Pa s \\
  $\nu$ \> kinematic viscosity, m$^2$/s \\
  $\xi_m$ \> coordinate in computational space \\
  $\rho$ \> density, kg/m$^3$ \\
  $\phi$ \> pressure correction value \\
  {\it Subscript} \\
  $i, j, k$ \> coordinate direction identifiers, \\
            \> or grid points \\
  $m, n$ \> coordinate direction identifiers \\
  w \> wall \\
  ref \> reference value \\
  {\it Superscript} \\
  $l$ \> number of simultaneous relaxation \\
  $m$ \> Newton iterative level \\
  $n$ \> time level \\
  $*$ \> non-dimensional variable \\
  $\hat{\, \, }$ \> predicted value \\
  $\overline{\, \, }$ \> interpolated value \\
\end{tabbing}
\end{multicols}
\end{framed}
\end{table}

\section{Fundamental equations}
\label{fundamental_equation}

This study deals with three-dimensional incompressible viscous flow 
and considers natural convection with small density change. 
The non-dimensionalized fundamental equations are the continuity equation, 
the Navier-Stokes equation under the Boussinesq approximation, 
and the energy equation, as follows:
\begin{equation}
   \frac{\partial u_j}{\partial x_j} = 0,
   \label{continuity}
\end{equation}
\begin{equation}
   \frac{\partial u_i}{\partial t} + \frac{\partial}{\partial x_j}(u_j u_i) 
   = - \frac{\partial p}{\partial x_i} 
   + \frac{1}{Re} \frac{\partial^2 u_i}{\partial x_j \partial x_j} 
   - \frac{Ra}{Re^2 Pr} T e_i,
   \label{navier-stokes}
\end{equation}
\begin{equation}
   \frac{\partial T}{\partial t} 
   + \frac{\partial}{\partial x_j}(u_j T) 
   = \frac{1}{Re Pr} \frac{\partial^2 T}{\partial x_j \partial x_j},
   \label{energy}
\end{equation}
where $i, j = 1, 2, 3$ are the $x$-, $y$-, and $z$-components, respectively. 
$t$ is the time, $x_i$ is coordinate, $u_i$ is the flow velocity, 
$p$ is the pressure, $T$ is the temperature, $g$ is the acceleration of gravity, 
$e_i$ is the unit vector in the direction of gravity. 
$Re$ is the Reynolds number, $Pr$ is the Prandtl number, 
and $Ra$ is the Rayleigh number. 
As reference values used for non-dimensionalization, 
the length is $l_\mathrm{ref}$, the velocity is $u_\mathrm{ref}$, 
and the temperature is $T_\mathrm{ref}$. 
The variables of the fundamental equations were non-dimensionalized 
by using these reference values as follows:
\begin{equation}
   x_i^{*} = \frac{x_i}{l_\mathrm{ref}}\ , \quad
   u_i^{*} = \frac{u_i}{u_\mathrm{ref}}\ , \quad
   p^{*} = \frac{p}{\rho u_\mathrm{ref}^{2}}\ , \quad
   t^{*} = \frac{u_\mathrm{ref} t}{l_\mathrm{ref}}, \quad
   T^{*} = \frac{T - T_\mathrm{ref}}{\Delta T}, \quad 
   \Delta T = T_\mathrm{w} - T_\mathrm{ref},
\end{equation}
where the superscript $*$ represents the non-dimensional variable, 
and the superscript $*$ was omitted in the fundamental equations. 
$T_\mathrm{w}$ is the temperature at the wall. 
The non-dimensional parameters in the fundamental equation are defined as
\begin{equation}
   Re = \frac{u_\mathrm{ref} l_\mathrm{ref}}{\nu}, \quad 
   Pr = \frac{\nu}{\alpha}, \quad 
   Ra = \frac{g \beta \Delta T l_\mathrm{ref}^3}{\nu \alpha},
\end{equation}
where $\nu$ is the kinematic viscosity of the fluid, 
$\alpha$ is the thermal diffusivity coefficient, 
and $\beta$ is the volume expansion coefficient.

Kinetic energy is defined as $K = u_i u_i/2$. 
For inviscid flow with $Re = \infty$, 
multiplying Eq. (\ref{navier-stokes}) by the velocity yields 
the following kinetic energy transport equation:
\begin{equation}
   \frac{\partial K}{\partial t} + \frac{\partial}{\partial x_j}(u_j K)
   = - \frac{\partial u_i p}{\partial x_i} + p \frac{\partial u_i}{\partial x_i}.
   \label{kinetic_energy_eq}
\end{equation}
From the continuity equation (\ref{continuity}), 
the second term on the right-hand side becomes zero, 
and the above equation (\ref{kinetic_energy_eq}) is in conservative form. 
Therefore, kinetic energy is conserved in inviscid periodic flows. 
Depending on the discretization of the continuity equation (\ref{continuity}) 
and the Navier--Stokes equation (\ref{navier-stokes}), 
the accuracy of the second term of Eq. (\ref{kinetic_energy_eq}) changes. 
Therefore, the discretization of the fundamental equation affects 
the conservation and calculation accuracy for kinetic energy.

\section{Coordinate transformation}
\label{transformation}

The fundamental equations in the Cartesian coordinate system are transformed 
into the general curvilinear coordinate system 
so that flow fields with arbitrary-shaped boundaries can be analyzed. 
The relationship between the coordinates $x_i$ in the physical space 
and the computational space $\xi_m$ is given as follows:
\begin{equation}
   x_i = x_i (\xi_1, \xi_2, \xi_3), \quad 
   \xi_m = \xi_m (x_1, x_2, x_3).
\end{equation}
By using the above relation, 
the fundamental equations are transformed into the general curvilinear coordinate system as follows:
\begin{equation}
   \frac{1}{J} \frac{\partial J U_m}{\partial \xi_m} = 0,
   \label{bfc.continuity}
\end{equation}
\begin{equation}
   \frac{\partial u_i}{\partial t} 
   + \frac{1}{J} \frac{\partial}{\partial \xi_m}(JU_m u_i)
   = - \frac{\partial \xi_m}{\partial x_i} \frac{\partial p}{\partial \xi_m}
   + \frac{1}{Re} \frac{1}{J} \frac{\partial}{\partial \xi_m}
     \left( Jg_{mn} \frac{\partial u_i}{\partial \xi_n} \right)
   - \frac{Ra}{Re^2 Pr} T e_i,
   \label{bfc.navier-stokes}
\end{equation}
\begin{equation}
   \frac{\partial T}{\partial t} 
   + \frac{1}{J} \frac{\partial}{\partial \xi_m}(JU_m T)
   = \frac{1}{Re Pr} \frac{1}{J} \frac{\partial}{\partial \xi_m}
     \left( Jg_{mn} \frac{\partial T}{\partial \xi_n} \right),
   \label{bfc.energy}
\end{equation}
where $J$ is the Jacobian, $U_m$ is the contravariant velocity component, 
and $g_{mn}$ is the metric tensor, which are defined by the following equations:
\begin{equation}
   J = \mbox{det} \biggl( \frac{\partial x_i}{\partial \xi_j} \biggl),
   \label{jacobian}
\end{equation}
\begin{equation}
   U_m = \frac{\partial \xi_m}{\partial x_j} u_j,
   \label{contravariant_velocity}
\end{equation}
\begin{equation}
   g_{mn} = \frac{\partial \xi_m}{\partial x_j} 
            \frac{\partial \xi_n}{\partial x_j}.
   \label{metrics_tensor}
\end{equation}
det in the definition of the Jacobian $J$ represents 
the determinant of the transformation matrix $[\partial x_i/\partial \xi_j]$. 
Each metric component $\partial \xi_m/\partial x_j$ and the Jacobian $J$ are given as
\begin{equation}
   \frac{\partial \xi_1}{\partial x_i} = \epsilon_{ijk} \frac{1}{J} 
   \frac{\partial x_j}{\partial \xi_2} \frac{\partial x_k}{\partial \xi_3}, \quad 
   \frac{\partial \xi_2}{\partial x_i} = \epsilon_{ijk} \frac{1}{J} 
   \frac{\partial x_j}{\partial \xi_3} \frac{\partial x_k}{\partial \xi_1}, \quad 
   \frac{\partial \xi_3}{\partial x_i} = \epsilon_{ijk} \frac{1}{J} 
   \frac{\partial x_j}{\partial \xi_1} \frac{\partial x_k}{\partial \xi_2},
\end{equation}
\begin{equation}
   J = \epsilon_{ijk} \frac{\partial x_i}{\partial \xi_1} 
                     \frac{\partial x_j}{\partial \xi_2} 
                     \frac{\partial x_k}{\partial \xi_3},
\end{equation}
where $\epsilon_{ijk}$ is the alternation symbol.

\section{Weighted interpolation method for cell interface velocity}
\label{interpolation}

\subsection{Definitions of finite difference and interpolation operations}

The variables at a cell center $(i, j, k)$ are defined as $\Phi_{i,j,k}$ and $\Psi_{i,j,k}$. 
The second-order central difference equation and interpolation 
for the variable $\Phi$ and the permanent product for two variables are as follows \citep{Morinishi_1998}:
\begin{equation}
   \left. \frac{\partial \Phi}{\partial x_1} \right|_{i,j,k} 
   = \frac{\delta_1 \Phi}{\delta_1 x_1} 
   = \frac{\bar{\Phi}^{x_1}_{i+1/2,j,k} - \bar{\Phi}^{x_1}_{i-1/2,j,k}}{\Delta x_1},
\end{equation}
\begin{equation}
   \left. \frac{\partial \Phi}{\partial x_1} \right|_{i,j,k} 
   = \frac{\delta_2 \Phi}{\delta_2 x_1} 
   = \frac{\Phi_{i+1,j,k} - \Phi_{i-1,j,k}}{2 \Delta x_1},
\end{equation}
\begin{equation}
   \bar{\Phi}^{x_1} = \frac{\Phi_{i,j,k} + \Phi_{i+1,j,k}}{2}.
\end{equation}
\begin{equation}
   \left. \widetilde{\Phi \Psi}^{x_1} \right|_{i+1/2,j,k} 
   = \frac{\Phi_{i,j,k} \Psi_{i+1,j,k} 
         + \Phi_{i+1,j,k} \Psi_{i,j,k}}{2},
\end{equation}
where $\Delta x_1$ is the grid width in the physical space. 
The difference formula and interpolation for the $x_2$- and $x_3$-directions are similarly defined. 
The difference formula and interpolation in the case of coordinate transformation are similarly defined and given as
\begin{equation}
   \left. \frac{1}{J} \frac{\partial J \xi_{1,1} \Phi}{\partial \xi_1} \right|_{i,j,k} 
      = \frac{1}{J} \frac{(J \xi_{1,1} \bar{\Phi}^{\xi_1})_{i+1/2,j,k} 
                        - (J \xi_{1,1} \bar{\Phi}^{\xi_1})_{i-1/2,j,k}}{\Delta \xi_1},
\end{equation}
\begin{equation}
   \left. \bar{\Phi}^{\xi_1} \right|_{i+1/2,j,k} 
   = \frac{\Phi_{i,j,k} + \Phi_{i+1,j,k}}{2},
\end{equation}
where $\Delta \xi_1$ is the grid width in the computational space. 
The difference formula and interpolation for the $\xi_2$- and $\xi_3$-directions are similarly defined. 
The Jacobian is defined at cell centers.

If a variable at time level $n$ is defined as $\Phi^n$, 
the derivative and interpolation for the variable in time direction are similarly given as
\begin{equation}
   \left. \frac{\partial \Phi}{\partial t} \right|^{n+1/2} 
      = \frac{\Phi^{n+1} - \Phi^{n}}{\Delta t},
\end{equation}
\begin{equation}
   \Phi^{n+1/2} = \frac{\Phi^{n+1} + \Phi^{n}}{2},
\end{equation}
where $\Delta t$ is a time increment.

\subsection{Interpolation of cell interface velocity}

We explain the accuracy of Rhie--Chow pressure interpolation \citep{Rhie&Chow_1983} 
for unsteady analysis. 
We will use the fundamental equations without buoyancy force 
in the Cartesian coordinate system to briefly explain the interpolation method. 
The derivation of the following equations is the same, 
even using the coordinate-transformed governing equations. 
Regarding the discretization of the continuity equation (\ref{continuity}) 
and the Navier--Stokes equation (\ref{navier-stokes}), 
we apply the implicit midpoint rule for the time derivative 
and the second-order central difference scheme for the spatial derivative. 
The discretized equation is given as
\begin{equation}
   \frac{\delta_2 u_i^{n+1,m+1}}{\delta_2 x_i} = 0,
   \label{continuity_mpr_u}
\end{equation}
\begin{equation}
   \frac{\delta_1 U_i^{n+1,m+1}}{\delta_1 x_i} = 0,
   \label{continuity_mpr_U}
\end{equation}
\begin{equation}
   \frac{u_i^{n+1,m+1} - u_i^n}{\Delta t} 
   = H_i^{n+\lambda,m+1} - \frac{\delta_2 p^{n+\lambda,m+1}}{\delta_2 x_i},
   \label{navier-stokes_mpr}
\end{equation}
\begin{equation}
   H_i^{n+\lambda,m+1} 
   = - \frac{\delta_1 U_j^{n+\lambda,m+1} \overline{u_i^{n+\lambda,m+1}}^{x_j}}{\delta_1 x_j} 
   + \frac{1}{Re} \frac{\delta_1}{\delta_1 x_j} 
   \left( \frac{\delta_1 u_i^{n+\lambda,m+1}}{\delta_1 x_j} \right),
\end{equation}
\begin{equation}
   u_i^{n+\lambda,m+1} = \lambda  u_i^{n+1,m+1} + (1 - \lambda) u_i^n,
\end{equation}
\begin{equation}
   U_i^{n+\lambda,m+1} = \lambda  U_i^{n+1,m+1} + (1 - \lambda) U_i^n,
\end{equation}
\begin{equation}
   p^{n+\lambda,m+1} = \lambda  p^{n+1,m+1} + (1 - \lambda) p^n,
\end{equation}
where $U_i$ represents the velocity at the cell interface. 
The superscripts $n$ and $m$ indicate the time and Newton iterative levels, respectively. 
The Newton method was applied to Eqs. (\ref{continuity_mpr_u}), 
(\ref{continuity_mpr_U}) and (\ref{navier-stokes_mpr}) to solve unsteady solutions. 
Regarding the discretization of the time derivative, 
if $\lambda = 1$, the Euler implicit method is applied to the time derivative. 
If $\lambda = 1/2$, the implicit midpoint rule is applied. 

Applying the simplified marker and cell (SMAC) method \citep{Amsden&Harlow_1970}, 
Eq. (\ref{navier-stokes_mpr}) is temporally split as follows:
\begin{equation}
   \frac{\hat{u}_i^{n+1,m+1} - u_i^n}{\Delta t} 
   = H_i^{n+\lambda,m+1} - \frac{\delta_2 p^{n+\lambda,m}}{\delta_2 x_i},
   \label{predict.mpr_u}
\end{equation}
\begin{equation}
   \frac{u_i^{n+1,m+1} - \hat{u}_i^{n+1,m+1}}{\Delta t} 
   = - \lambda \frac{\delta_2 \phi^m}{\delta_2 x_i},
   \label{correct.mpr_u}
\end{equation}
\begin{equation}
   p^{n+1,m+1} = p^{n+1,m} + \phi^m,
\end{equation}
where $\hat{u}_i^{n+1,m+1}$ is the predicted value of velocity, 
and $\phi^m$ is the pressure correction value. 
The velocity in $H_i^{n+1,m+1}$ is defined as 
$u_i^{n+\lambda} = \lambda  \hat{u}_i^{n+1} + (1 - \lambda) u_i^n$. 
Taking the divergence of Eq. (\ref{correct.mpr_u}) and 
using the continuity equation (\ref{continuity_mpr_u}) at the $n+1$ level, 
the Poisson equation for the pressure correction value $\phi^m$ is derived as
\begin{equation}
   \lambda \frac{\delta_2}{\delta_2 x_i} 
   \left( \frac{\delta_2 \phi^m}{\delta_2 x_i} \right) 
   = \frac{1}{\dt} \frac{\delta_2 \hat{u}_i^{n+1,m+1}}{\delta_2 x_i}.
   \label{poisson_mpr_u}
\end{equation}
In existing studies 
\citep{Rhie&Chow_1983,Peric_et_al_1988,Majumdar_1988,Zang_et_al_1994,Morinishi_1998,Choi_1999,Morinishi_1999,Yu_et_al_2002,Bartholomew_et_al_2018,Lee_et_al_2019}, 
the formula (\ref{poisson_mpr_u}) was not used to obtain pressure or pressure correction value. 
The equation for modifying the velocity $U_i$ at the cell interface is given as
\begin{equation}
   \frac{U_i^{n+1,m+1} - \hat{U}_i^{n+1,m+1}}{\Delta t} 
   = - \lambda \frac{\delta_1 \phi^m}{\delta_1 x_i}.
   \label{correct.cellface.mpr_u}
\end{equation}
Taking the divergence of Eq. (\ref{correct.cellface.mpr_u}) 
and using the continuity equation (\ref{continuity_mpr_U}) at the $n+1$ level, 
the Poisson equation for the pressure correction value $\phi^m$ is derived as
\begin{equation}
   \lambda \frac{\delta_1}{\delta_1 x} 
   \left( \frac{\delta_1 \phi^m}{\delta_1 x_i} \right) 
   = \frac{1}{\Delta t} \frac{\delta_1 \hat{U}_i^{n+1,m+1}}{\delta_1 x_i}.
   \label{poisson_mpr_U}
\end{equation}
Generally, Eq. (\ref{poisson_mpr_U}) is used to solve the pressure or pressure correction value. 
Using the velocity $\hat{u}_i^{n+1,m+1}$ at the cell center, 
the velocity $\hat{U}_i^{n+1,m+1}$ at the cell interface is obtained 
by direct interpolation as follows:
\begin{equation}
   \hat{U}_i^{n+1,m+1} = \overline{\hat{u}_i^{n+1,m+1}}^{x_i}.
   \label{cellface.mpr_u.0}
\end{equation}
The continuity equation $\delta_2 u_i/\delta_2 x_i$ determined by the velocity 
at the cell center affects kinetic energy. 
Therefore, similar to existing research \citep{Morinishi_1999}, 
we evaluate the error of the continuity formula. 
Using Eqs. (\ref{correct.mpr_u}) and (\ref{correct.cellface.mpr_u}), 
the continuity equation is given as
\begin{eqnarray}
   \frac{\delta_2 u_i^{n+1,m+1}}{\delta_2 x_i} 
   &=& \frac{\delta_1 \overline{u_i^{n+1,m+1}}^{x_i}}{\delta_1 x_i} 
   = \frac{\delta_1}{\delta_1 x_i} \left( 
   \overline{\hat{u}_i^{n+1,m+1}}^{x_i} - \Delta t \lambda \overline{\frac{\delta_2 \phi^m}{\delta_2 x_i}}^{x_i} 
   \right) 
   \nonumber \\
   &=& \frac{\delta_1}{\delta_1 x_i} \left( 
   U_i^{n+1,m+1} + \Delta t \lambda \frac{\delta_1 \phi^m}{\delta_1 x_i} 
   - \Delta t \lambda \overline{\frac{\delta_2 \phi^m}{\delta_2 x_i}}^{x_i} 
   \right) 
   \nonumber \\
   &=& \frac{\delta_1 U_i^{n+1,m+1}}{\delta_1 x_i} 
   + \Delta t \lambda \frac{\delta_1}{\delta_1 x_i} \left( 
   \frac{\delta_1 \phi^m}{\delta_1 x_i} 
   - \overline{\frac{\delta_2 \phi^m}{\delta_2 x_i}}^{x_i} 
   \right).
   \label{continuity_mpr_fdm1}
\end{eqnarray}
If the continuity equation $\delta_1 U_i/\delta_1 x_i = 0$ obtained 
using the cell interface velocities is satisfied discretely, 
then the first term in the formula (\ref{continuity_mpr_fdm1}) can be ignored. 
Expanding the discrete value of the pressure correction value into a Taylor series, 
we can evaluate the error of the formula (\ref{continuity_mpr_fdm1}) as follows:
\begin{eqnarray}
   \frac{\delta_2 u_i^{n+1,m+1}}{\delta_2 x_i} 
   &=& \Delta t \frac{\delta_1}{\delta_1 x_i} \left( 
   \frac{\delta_1 \phi^m}{\delta_1 x_i} 
   - \overline{\frac{\delta_2 \phi^m}{\delta_2 x_i}}^{x_i} 
   \right) 
   \sim 
   - \frac{\Delta t}{4} \frac{\partial^4 \phi}{\partial x_i^4} \dx_i^2 
   = O(\Delta t^2 \dx_i^2),
\end{eqnarray}
where $\phi = O(\Delta t)$ is used. 
The error of the continuity equation $\delta_2 u_i^{n+1,m+1}/\delta_2 x_i = 0$ 
is the second-order accuracy in time and space.

When using Eq. (\ref{cellface.mpr_u.0}), oscillation may occur in pressure distribution. 
To avoid spurious errors of pressure, 
using Rhie--Chow pressure interpolation \citep{Rhie&Chow_1983}, 
the velocity $\hat{U}_i^{n+1,m+1}$ at the cell interface is given as
\begin{equation}
   \hat{U}_i^{n+1,m+1} 
   = \overline{\hat{u}_i^{n+1,m+1}}^{x_i} 
   + \Delta t \lambda \overline{ \frac{\delta_2 p^{n+1,m}}{\delta_2 x_i} }^{x_i} 
   - \Delta t \lambda \frac{\delta_1 p^{n+1,m}}{\delta_1 x_i}.
   \label{cellface.mpr_u.1}
\end{equation}
When the pressure at the cell center $(i, j, k)$ is expanded into a Taylor series, 
Eq. (\ref{cellface.mpr_u.1} ) can be rewritten as follows:
\begin{equation}
   \hat{U}_i^{n+1,m+1} 
   = \overline{\hat{u}_i^{n+1,m+1}}^{x_i} 
   + \Delta t \frac{\dx_i^2}{4} 
     \frac{\partial^3 p^{n+1,m}}{\partial x_i^3} + O(\dx^4).
\end{equation}
The cell interface velocities contain errors of the first order in time 
and second order in space.

Using Eqs. (\ref{correct.mpr_u}), (\ref{correct.cellface.mpr_u}), and (\ref{cellface.mpr_u.1}), 
the continuity equation obtained by the cell center velocity is given as
\begin{eqnarray}
   \frac{\delta_2 u_i^{n+1,m+1}}{\delta_2 x_i} 
   &=& \frac{\delta_1 \overline{u_i^{n+1,m+1}}^{x_i}}{\delta_1 x_i} 
   = \frac{\delta_1}{\delta_1 x_i} \left( 
   \overline{\hat{u}_i^{n+1,m+1}}^{x_i} - \Delta t \lambda \overline{\frac{\delta_2 \phi^m}{\delta_2 x_i}}^{x_i} 
   \right) 
   \nonumber \\
   &=& \frac{\delta_1}{\delta_1 x_i} \left( 
   U_i^{n+1,m+1} 
   - \Delta t \lambda \overline{\frac{\delta_2 p^{n+1,m}}{\delta_2 x_i}}^{x_i} 
   + \Delta t \lambda \frac{\delta_1 p^{n+1,m}}{\delta_1 x_i} 
   + \Delta t \lambda \frac{\delta_1 \phi^m}{\delta_1 x_i} 
   - \Delta t \lambda \overline{\frac{\delta_2 \phi^m}{\delta_2 x_i}}^{x_i} 
   \right) 
   \nonumber \\
   &=& \frac{\delta_1 U_i^{n+1,m+1}}{\delta_1 x_i} 
   + \Delta t \lambda \frac{\delta_1}{\delta_1 x_i} \left( 
   \frac{\delta_1 p^{n+1,m+1}}{\delta_1 x_i} 
   - \overline{\frac{\delta_2 p^{n+1,m+1}}{\delta_2 x_i}}^{x_i} 
   \right) 
   \nonumber \\
   &\sim& 
   - \frac{\Delta t}{4} \frac{\partial^4 p}{\partial x_i^4} \dx_i^2 
   = O(\Delta t^1 \dx_i^2).
   \label{continuity_mpr_fdm2}
\end{eqnarray}
The error of the continuity equation $\delta_2 u_i^{n+1,m+1}/\delta_2 x_i = 0$ 
is the first order accuracy for time and the second order for space.

We improve the interpolation of the velocity at a cell interface 
so that the errors concerning time are second-order accurate in the continuity equation 
$\delta_2 u_i^{n+1}/\delta_2$. 
The expression (\ref{predict.mpr_u}) is rewritten as
\begin{equation}
   \frac{\hat{u}_i^{n+1,m+1} - u_i^n}{\Delta t} 
   = H_i^{n+\lambda,m+1} - \lambda \frac{\delta_2 (p^{n+1,m} - p^n)}{\delta_2 x_i} 
   - \frac{\delta_2 p^{n}}{\delta_2 x_i}.
\end{equation}
To find the velocity at the cell interface, 
we use the second term on the right-hand side of the above equation. 
The velocity $\hat{U}_i^{n+1,m+1}$ at the cell interface is obtained 
using the pressure difference $\Delta p^{n+1,m} = p^{n+1,m} - p^n$ as follows:
\begin{equation}
   \hat{U}_i^{n+1,m+1} = \overline{\hat{u}_i^{n+1,m+1}}^{x_i} 
   + \Delta t \lambda \overline{ \frac{\delta_2 \Delta p^{n+1,m}}{\delta_2 x_i} }^{x_i} 
   - \Delta t \lambda \frac{\delta_1 \Delta p^{n+1,m}}{\delta_1 x_i}.
   \label{cellface.mpr_u.2}
\end{equation}
This pressure difference weighted interpolation was also used 
in the study of \citet{Lee_et_al_2019}. 
Expanding the pressure difference $\Delta p$ to a Taylor series, 
Eq. (\ref{cellface.mpr_u.2}) is given as
\begin{equation}
   \hat{U}_i^{n+1,m+1} = \overline{\hat{u}_i^{n+1,m+1}}^{x_i} 
   + \Delta t \frac{\dx_i^2}{4} \frac{\partial^3 \Delta p^{n+1,m}}{\partial x^3} + O(\dx^4).
\end{equation}
Because of $\Delta p^{n+1,m} = p^{n+1,m}-p^n = O(\Delta t)$, 
the error in the cell interface velocity is the second-order accuracy 
for both time and space. 
The method using Rhie--Chow pressure interpolation \citep{Rhie&Chow_1983} 
shown in Eq. (\ref{cellface.mpr_u.1}) contains errors of first-order accuracy in time. 
However, using the Eq. (\ref{cellface.mpr_u.2}), 
the time discretization is second-order accuracy. 
In this research, Newton iteration is not performed when obtaining steady fields. 
Therefore, we replace $\Delta p^{n+1,m}$ in Eq. (\ref{cellface.mpr_u.2}) with $p^{n+1,m}$. 
In other words, Eq. (\ref{cellface.mpr_u.1}) is used to interpolate the cell interface velocity.

Next, we evaluate the error of the continuity equation 
$\delta_2 u_i/\delta_2 x_i = 0$ determined by the velocity at the cell center. 
Using Eqs. (\ref{correct.mpr_u}), (\ref{correct.cellface.mpr_u}), 
and (\ref{cellface.mpr_u.2}), the continuity formula is given as
\begin{eqnarray}
   \frac{\delta_2 u_i^{n+1,m+1}}{\delta_2 x_i} 
   &=& \frac{\delta_1 \overline{u_i^{n+1,m+1}}^{x_i}}{\delta_1 x_i} 
   = \frac{\delta_1}{\delta_1 x_i} \left( 
   \overline{\hat{u}_i^{n+1,m+1}}^{x_i} - \Delta t \lambda \overline{\frac{\delta_2 \phi^m}{\delta_2 x_i}}^{x_i} 
   \right) 
   \nonumber \\
   &=& \frac{\delta_1}{\delta_1 x_i} \left( 
   U_i^{n+1,m+1} 
   - \Delta t \lambda \overline{\frac{\delta_2 \Delta p^{n+1,m}}{\delta_2 x_i}}^{x_i} 
   + \Delta t \lambda \frac{\delta_1 \Delta p^{n+1,m}}{\delta_1 x_i} 
   + \Delta t \lambda \frac{\delta_1 \phi^m}{\delta_1 x_i} 
   - \Delta t \lambda \overline{\frac{\delta_2 \phi^m}{\delta_2 x_i}}^{x_i} 
   \right) 
   \nonumber \\
   &=& \frac{\delta_1 U_i^{n+1,m+1}}{\delta_1 x_i} 
   + \Delta t \lambda \frac{\delta_1}{\delta_1 x_i} \left( 
   \frac{\delta_1 (p^{n+1,m+1} - p^n)}{\delta_1 x_i} 
   - \overline{\frac{\delta_2 (p^{n+1,m+1} - p^n)}{\delta_2 x_i}}^{x_i} 
   \right) 
   \nonumber \\
   &\sim& 
   - \frac{\Delta t}{4} \frac{\partial^4 \Delta p}{\partial x_i^4} \dx_i^2 
   = O(\Delta t^2 \dx_i^2),
   \label{continuity_mpr_fdm3}
\end{eqnarray}
where $\Delta p^{n+1} = p^{n+1} - p^n = O(\Delta t)$ 
after the Newton iteration finishes. 
The error of the continuity formula $\delta_2 u_i^{n+1}/\delta_2 x_i = 0$ 
is the second-order accuracy in both time and space.

In the collocation method, the velocity and pressure are obtained 
so as to satisfy the continuity equation (\ref{continuity_mpr_U}) 
obtained from the velocity at the cell interface. 
Therefore, the continuity equation (\ref{continuity_mpr_u}) 
obtained from the velocity at the cell center is not satisfied. 
Here, we consider a method that satisfies the two continuity equations. 
The pressure $p$ is calculated from the Poisson equation (\ref{poisson_mpr_u}) 
so as to satisfy the continuity equation (\ref{continuity_mpr_u}). 
Furthermore, the pressure $p_f$ is obtained from the Poisson equation (\ref{poisson_mpr_U}) 
so as to satisfy the continuity equation (\ref{continuity_mpr_U}). 
Similar to Rhie--Chow pressure interpolation \citep{Rhie&Chow_1983}, 
we define the velocity $\hat{U}_i^{n+1,m+1}$ at the cell interface as follows:
\begin{equation}
   \hat{U}_i^{n+1,m+1} 
   = \overline{\hat{u}_i^{n+1,m+1}}^{x_i} 
   + \dt \lambda \overline{ \frac{\delta_2 p^{n+1,m}}{\delta_2 x_i} }^{x_i} 
   - \dt \lambda \frac{\delta_1 p_f^{n+1,m}}{\delta_1 x_i}.
   \label{cellface.u_3.1}
\end{equation}
Two pressures are used in the above equation to interpolate the velocity at the cell interface. 
Expanding the pressure at the cell center $(i, j, k)$ into a Taylor series, 
Eq. (\ref{cellface.u_3.1}) can be rewritten as
\begin{equation}
   \hat{U}_i^{n+1,m+1} 
   = \overline{\hat{u}_i^{n+1,m+1}}^{x_i} 
   + \dt \left( \frac{7 \dx_i^2}{24} 
     \frac{\partial^3 p^{n+1,m}}{\partial x_i^3} 
   + \frac{\partial p^{n+1,m}}{\partial x_i} \right) 
   - \dt \left( \frac{\dx_i^2}{24} 
     \frac{\partial^3 p_f^{n+1,m}}{\partial x_i^3} 
   + \frac{\partial p_f^{n+1,m}}{\partial x_i} \right) + O(\dx^4).
\end{equation}
The error includes $\partial p^{n+1,m}/\partial x_i$ and $\partial p_f^{n+1,m}/\partial x_i$. If $p=p_f$, the error terms of these first derivatives cancel each other out. Then, the velocity at the cell interface contains the errors of the first order in time and the second order in space.

Using Eqs. (\ref{correct.mpr_u}), (\ref{correct.cellface.mpr_u}), and (\ref{cellface.u_3.1}), 
the continuity equation obtained by the cell center velocity is given as
\begin{eqnarray}
   \frac{\delta_2 u_i^{n+1,m+1}}{\delta_2 x_i} 
   &=& \frac{\delta_1 \overline{u_i^{n+1,m+1}}^{x_i}}{\delta_1 x_i} 
   = \frac{\delta_1}{\delta_1 x_i} \left( 
   \overline{\hat{u}_i^{n+1,m+1}}^{x_i} - \dt \lambda \overline{\frac{\delta_2 \phi^m}{\delta_2 x_i}}^{x_i} 
   \right) 
   \nonumber \\
   &=& \frac{\delta_1}{\delta_1 x_i} \left( 
   U_i^{n+1,m+1} 
   - \dt \lambda \overline{\frac{\delta_2 p^{n+1,m}}{\delta_2 x_i}}^{x_i} 
   + \dt \lambda \frac{\delta_1 p_f^{n+1,m}}{\delta_1 x_i} 
   + \dt \lambda \frac{\delta_1 \phi^m}{\delta_1 x_i} 
   - \dt \lambda \overline{\frac{\delta_2 \phi^m}{\delta_2 x_i}}^{x_i} 
   \right) 
   \nonumber \\
   &=& \frac{\delta_1 U_i^{n+1,m+1}}{\delta_1 x_i} 
   + \dt \lambda \frac{\delta_1}{\delta_1 x_i} \left( 
   \frac{\delta_1 p_f^{n+1,m+1}}{\delta_1 x_i} 
   - \overline{\frac{\delta_2 p^{n+1,m+1}}{\delta_2 x_i}}^{x_i} 
   \right) 
   \nonumber \\
   &\sim& 
     \dt \left( \frac{1}{12} \frac{\partial^4 p_f}{\partial x_i^4} \dx_i^2 
   + \frac{\partial^2 p_f}{\partial x_i^2} \right) 
   - \dt \left( \frac{1}{3} \frac{\partial^4 p}{\partial x_i^4} \dx_i^2 
   + \frac{\partial^2 p}{\partial x_i^2} \right).
   \label{continuity3_fdm1}
\end{eqnarray}
For $p = p_f$, the continuity formula (\ref{continuity3_fdm1}) agrees with Eq. (\ref{continuity_mpr_fdm2}). 
Then the continuity formula is $\delta_2 u_i^{n+1,m+1}/\delta_2 x_i = O(\dt^1 \dx_i^2)$, 
and the error is the first-order accuracy for time 
and the second-order accuracy for space.

We refine the interpolation of the cell interface velocity 
so that the error in the continuity equation 
$\delta_2 u_i^{n+1}/\delta_2 x_i = 0$ to time is second-order accurate. 
We replace the pressures $p$ and $p_f$ with the pressure differences $\Delta p$ and $\Delta p_f$ 
in Eq. (\ref{cellface.u_3.1}). 
The velocity $\hat{U}_i^{n+1,m+1}$ at the cell interface is given as
\begin{equation}
   \hat{U}_i^{n+1,m+1} 
   = \overline{\hat{u}_i^{n+1,m+1}}^{x_i} 
   + \dt \lambda \overline{ \frac{\delta_2 \Delta p^{n+1,m}}{\delta_2 x_i} }^{x_i} 
   - \dt \lambda \frac{\delta_1 \Delta p_f^{n+1,m}}{\delta_1 x_i}.
   \label{cellface.u_3.2}
\end{equation}
It is found that using the Taylor series expansion, 
the velocity at the cell interface contains the errors of the second order in time. 
In addition, the error in the continuity formula 
$\delta_2 u_i^{n+1}/\delta_2 x_i = 0$ given by the velocity at the cell center 
is the second-order accuracy in time.

We should calculate two Poisson equations to obtain two pressures 
for the pressure interpolation using Eq. (\ref{cellface.u_3.1}). 
Therefore, it leads to an increase in computation time. 
As a result of numerical experiments, 
when $p \approx p_f$, two continuity equations (\ref{continuity_mpr_u}) 
and (\ref{continuity_mpr_U}) could be satisfied at the same time. 
However, if we changed the Courant number, 
$p \approx p_f$ did not hold, and no convergent solution was obtained. 
At present, we have not established a method that simultaneously satisfies 
the two continuity equations. 
Further investigation is necessary in the future.

\section{Numerical method}
\label{numerical_method}

An overview of the method of solving Eqs. (\ref{continuity}) and (\ref{navier-stokes}) 
in the Cartesian coordinate system was given in Section \ref{interpolation}. 
Equations (\ref{bfc.navier-stokes}) and (\ref{bfc.energy}) 
transformed to general curvilinear coordinates are solved similarly. 
Herein, we describe a method for the simultaneous relaxation of velocity and pressure. 
The technique is the same as the simultaneous relaxation method 
used by the authors \citep{Yanaoka&Inafune_2023, Yanaoka_2023}, 
and the existing process is extended to the general curvilinear coordinate system. 
The Newton method is used to solve the unsteady solution. 
Appling the implicit midpoint rule to Eqs. (\ref{bfc.navier-stokes}) 
and (\ref{bfc.energy}), the respective equations are given as
\begin{equation}
   \frac{u_i^{n+1,m+1} - u_i^n}{\Delta t} = H_i^{n+\lambda,m+1} 
   - \frac{\partial \xi_m}{\partial x_i} 
     \frac{\partial p^{n+\lambda,m+1}}{\partial \xi_m},
   \label{implicit.newton.u}
\end{equation}
\begin{equation}
   \frac{T^{n+1,m+1} - T^n}{\Delta t} = H_{T}^{n+\lambda,m+1},
   \label{implicit.newton.t}
\end{equation}
\begin{equation}
   H_i^{n+\lambda,m+1} 
   = - \frac{1}{J} \frac{\partial}{\partial \xi_m} (JU_m^{n+\lambda,m+1} u_i^{n+\lambda,m+1}) 
   + \frac{1}{Re} \frac{1}{J} \frac{\partial}{\partial \xi_m} 
     \left( Jg_{mn} \frac{\partial u_i^{n+\lambda,m+1}}{\partial \xi_n} \right)
   - \frac{Ra}{Re^2 Pr} T^{n+\lambda,m+1} e_i,
\end{equation}
\begin{equation}
   H_T^{n+\lambda,m+1} 
   = - \frac{1}{J} \frac{\partial}{\partial \xi_m}(JU_m^{n+\lambda,m+1} T^{n+\lambda,m+1}) 
   + \frac{1}{Re Pr} \frac{\partial}{\partial \xi_m} 
     \left( Jg_{mn} \frac{\partial T^{n+\lambda,m+1}}{\partial \xi_n} \right),
\end{equation}
\begin{equation}
   u_i^{n+\lambda,m+1} = \lambda  u_i^{n+1,m+1} + (1 - \lambda) u_i^n,
\end{equation}
\begin{equation}
   JU_m^{n+\lambda,m+1} = \lambda  JU_m^{n+1,m+1} + (1 - \lambda) JU_m^n,
\end{equation}
\begin{equation}
   p^{n+\lambda,m+1} = \lambda  p^{n+1,m+1} + (1 - \lambda) p^n,
\end{equation}
\begin{equation}
   T^{n+\lambda,m+1} = \lambda  T^{n+1,m+1} + (1 - \lambda) T^n,
\end{equation}
where the superscripts $n$ and $m$ indicate the time and Newton iterative levels, respectively. 
This study uses the Euler implicit method ($\lambda = 1$) for steady field analysis 
and the implicit midpoint rule ($\lambda = 1/2$) for unsteady field analysis. 
Spatial derivatives are discretized with second-order accuracy central differences 
using the method described in Section \ref{interpolation}.

We use a collocated grid in a general curvilinear coordinate system. 
Applying the SMAC method \citep{Amsden&Harlow_1970}, 
Eq. (\ref{implicit.newton.u}) is temporally split as follows:
\begin{equation}
   \frac{\hat{u}_i^{n+1,m+1} - u_i^n}{\Delta t} = H_i^{n+\lambda,m+1} 
   - \frac{\partial \xi_m}{\partial x_i} \frac{\partial}{\partial \xi_m} 
   \left[ \lambda p^{n+1,m} + (1 - \lambda) p^n \right],
   \label{implicit.predict.u}
\end{equation}
\begin{equation}
   \frac{u_i^{n+1,m+1} - \hat{u}_i^{n+1,m+1}}{\Delta t} = 
   - \lambda \frac{\partial \xi_m}{\partial x_i} 
             \frac{\partial \phi^m}{\partial \xi_m},
   \label{implicit.correct.u}
\end{equation}
\begin{equation}
   p^{n+1,m+1} = p^{n+1,m} + \phi^m,
   \label{implicit.correct.p}
\end{equation}
where $\hat{u}_i^{n+1,m+1}$ is the predicted value of velocity, 
and $\phi^m$ is the pressure correction value. 

Next, we must obtain the contravariant velocity $JU_m^{n+1}$ at the cell interface 
to derive the Poisson equation for the pressure correction value. 
From the contravariant velocity definition equation (\ref{contravariant_velocity}) 
and the velocity correction equation (\ref{implicit.correct.u}), 
the following equation is obtained:
\begin{equation}
   JU_m^{n+1,m+1} = \hat{JU}_m^{n+1,m+1} 
   - \Delta t \lambda \left( Jg_{mn} \frac{\partial \phi^m}{\partial \xi_n} \right).
   \label{implicit.correct.JU}
\end{equation}
As the predicted velocity value $\hat{u}_i$ is defined at the cell center, 
we must find $\hat{u}_i$ at the cell interface by interpolation 
to obtain $\hat{JU}_m$. 
Using the pressure interpolation by Rhie--Chow \citep{Rhie&Chow_1983}, 
the velocity at the cell interface can be defined as
\begin{equation}
   \hat{JU}_m^{n+1,m+1} = \left< JU_m^{n+1,m+1} \right> 
   - \Delta t \lambda Jg_{mm} 
   \left( \frac{\partial p^{n+1,m}}{\partial \xi_m} 
   - \left \langle \frac{\partial p^{n+1,m}}{\partial \xi_m} \right \rangle 
   \right),
   \label{implicit.cellface.velocity.2}
\end{equation}
where $\langle \,\,\rangle$ represents the interpolated value at the cell interface.

In this study, as explained in Section \ref{interpolation}, 
we Interpolate the velocity at the cell interface using the pressure difference 
$\Delta p^{n+1,m} = p^{n+1,m}-p^n$ as follows:
\begin{equation}
   \hat{JU}_m^{n+1,m+1} = \left< JU_m^{n+1,m+1} \right> 
   - \Delta t \lambda Jg_{mm} 
   \left( \frac{\partial \Delta p^{n+1,m}}{\partial \xi_m} 
   - \left \langle \frac{\partial \Delta p^{n+1,m}}{\partial \xi_m} \right \rangle 
   \right).
   \label{implicit.cellface.velocity.3}
\end{equation}

Taking the divergence of Eq. (\ref{implicit.correct.JU}) 
and using the continuity equation (\ref{bfc.continuity}) at the $n+1$ level, 
the following Poisson equation for the pressure correction value $\phi$ is derived as
\begin{equation}
   \lambda \frac{\partial}{\partial \xi_m} \left( 
   Jg_{mn} \frac{\partial \phi^m}{\partial \xi_n} \right)
   = \frac{1}{\Delta t} 
   \frac{\partial \hat{JU}_m^{n+1,m+1}}{\partial \xi_m}.
   \label{implicit.poisson.1}
\end{equation}
Equations (\ref{implicit.predict.u}), (\ref{implicit.correct.u}), 
(\ref{implicit.correct.p}), (\ref{implicit.cellface.velocity.3}), 
and (\ref{implicit.poisson.1}) are used 
when the collocation method is applied to the SMAC method \citep{Amsden&Harlow_1970} 
in a general curvilinear coordinate system. 
When the Poisson equation in the above formula (\ref{implicit.poisson.1}) is discretized, 
coefficients of nine components appear in a two-dimensional case 
and 19 components in a three-dimensional case. 
Therefore, it takes much time to iterate the Poisson equation. 
In this study, to reduce memory usage and simplify the calculation of the Poisson equation, 
we omit the differential term in which the cross term $Jg_{mn}$ appears. 
Simply omitting the cross term will require under-relaxation 
when solving the Poisson equation using an iterative method 
such as the successive over-relaxation (SOR) method. 
Therefore, in this study, 
we adopt the idea of the highly simplified marker and cell (HSMAC) method \citep{Hirt_et_al_1975} 
and perform simultaneous relaxation of velocity and pressure 
to prevent under-relaxation. 
We simplify the Poisson equation for pressure correction value as follows:
\begin{equation}
   \lambda \frac{\partial}{\partial \xi_m} \left( 
   Jg_{mm} \frac{\partial \phi^{m,l}}{\partial \xi_m} \right)
   = \frac{1}{\Delta t} 
   \frac{\partial JU_m^{n+1,m+1,l}}{\partial \xi_m}.
   \label{implicit.poisson.2}
\end{equation}
Velocity and pressure are modified as follows: 
Note that the cross term in the velocity correction equation is not omitted:
\begin{equation}
   JU_m^{n+1,m+1,l+1} = JU_m^{n+1,m+1,l} 
   - \Delta t \lambda \left( Jg_{mn} \frac{\partial \phi^{m,l}}{\partial \xi_n} \right),
   \label{implicit.correct.JU.2}
\end{equation}
\begin{equation}
   u_i^{n+1,m+1,l+1} = u_i^{n+1,m+1,l} 
   - \Delta t \lambda \left( \frac{\partial \xi_m}{\partial x_i}
                         \frac{\partial \phi^{m,l}}{\partial \xi_m} \right),
   \label{implicit.correct.2}
\end{equation}
\begin{equation}
   p^{n+1,m+1,l+1} = p^{n+1,m+1,l} + \phi^{m,l},
   \label{implicit.pressure.2}
\end{equation}
where the superscript $l$ is the number of iterations. 
When $l = 1$, let $JU_m^{n+1,m+1,l} = \hat{JU}_m^{n+1,m+1}$, $u_i^{n+1,m+1,l} = \hat{u}_i^{n+1,m+1}$, and $p^{n+1,m+1,l} = p^{n+1,m+1}$, 
the velocity and pressure are simultaneously relaxed. 
We repeat the calculation up to a predetermined iteration number. 
After the simultaneous relaxation is completed, 
we let $JU_m^{n+1,m+1} = JU_m^{n+1,m+1,l+1}$, $u_i^{n+1,m+1} = u_i^{n+1,m+1,l+1}$, and $p^{n+1,m+1} = p^{n+1,m+1,l+1}$. 
Takemitsu \citep{Takemitsu_1985} proposed a similar method 
that simultaneously iterates the velocity correction equation 
and the Poisson equation of the pressure correction. 
However, the Poisson equation for pressure should be solved 
after correcting the velocity. 
The present numerical method does not require solving the Poisson equation for the pressure. 
It is significant to include the cross term in the velocity correction equation (\ref{implicit.correct.2}). 
By simultaneously relaxing velocity and pressure, 
the influence of the cross term is considered for the velocity and pressure, 
and the velocity is corrected to satisfy the continuity equation. 
With such simultaneous relaxation, 
the Poisson equation can be solved without under-relaxation. 
However, the use of the SOR method is inconvenient 
because the optimal value of the acceleration relaxation coefficient changes 
depending on the number of grid points and the flow field. 
In this study, we used the biconjugate gradient stabilized method \citep{Vorst_1992} 
to solve simultaneous linear equations.

A boundary condition is required when solving the Poisson equation (\ref{implicit.poisson.2}) 
for pressure correction value. 
In this study, as the velocity and pressure are simultaneously relaxed 
while solving the Poisson equation, 
the boundary condition of the pressure correction value is simplified. 
If $JU_m^{n+1,l+1} = JU_m^{n+1,l}$ at the boundary, 
then $Jg_{mn} \partial \phi^{m,l}/\partial \xi_n = 0$ is obtained. 
Considering that the pressure correction value $\phi$ asymptotically approaches zero 
with iteration, and omitting the influence of the cross term, 
the condition for first derivative zero, $\partial \phi^{m,l}/\partial \xi_n = 0$, 
at the boundary is obtained. 
Because $\phi$ asymptotically approaches zero with iteration, 
the effect of this approximation on the inside of the computational domain is considered to be small.

These discretized equations are solved following the next procedure.
\begin{enumerate}[Step 1:]
\setlength{\leftskip}{25pt}
\setlength{\itemsep}{0pt}
\setlength{\parskip}{0pt}
\setlength{\labelsep}{5pt}
\item At $m = 1$, let $u_i^{n+1,m} = u_i^n$, $p^{n+1,m} = p^n$, and $T^{n+1,m} = T^n$.
\item Solve Eq. (\ref{implicit.predict.u}), 
      and predict the velocity $\hat{u}_i^{n+1,m+1}$.
\item Interpolate the cell interface velocity $\hat{JU}_m^{n+1,m+1}$ 
      from Eq. (\ref{implicit.cellface.velocity.3}).
\item The Poisson equation (\ref{implicit.poisson.2}) for pressure correction value $\phi^m$, 
      and velocity and pressure correction equations (\ref{implicit.correct.JU.2}), 
      (\ref{implicit.correct.2}), (\ref{implicit.pressure.2}) are simultaneously relaxed. 
      At the end of simultaneous relaxation, 
      set $JU_m^{n+1,m+1}$, $u_i^{n+1,m+1}$ , and $p^{n+1,m+1}$.
\item Solve Eq. (\ref{implicit.newton.t}) and find the temperature $T^{n+1,m+1}$.
\item Repeat Steps 2 to 5. 
      After the Newton iteration is completed, set $u_i^{n+1} = u_i^{n+1,m+1}$, 
      $p^{n+1} = p^{n+1,m+1}$, and $T^{n+1} = T^{n+1,m+1}$.
\item Advance the time step and return to Step 1.
\end{enumerate}

\section{Verification of numerical method}
\label{verification}

This research first analyzes steady fields and verifies the validity of this numerical method. In addition, we confirm that pressure oscillation does not occur. Next, we analyze unsteady fields and investigate the influence of weighted interpolation by pressure difference on the calculation accuracy.

\subsection{Natural convection inside a cavity}

We analyze natural convection in a square cavity 
and compare our results with existing results. 
The origin is placed at the bottom of the container, 
the $x$-and $y$-axes are in the horizontal and vertical directions, respectively, 
and the $z$-axis is perpendicular to the plane of the paper. 
The length of one side of the container is $H$, 
and all boundaries are surrounded by walls. 
A non-slip boundary condition is applied to the wall surface. 
The left and right wall surfaces are heated and cooled 
at uniform temperatures $T_H$ and $T_C$, respectively. 
Adiabatic conditions are imposed on the upper and lower wall surfaces. 
The pressure is obtained by second-order accuracy extrapolation. 
Periodic boundaries are imposed in the $z$-direction 
for the velocity, pressure, and temperature. 
The grid used is a $N \times N \times 2$ non-uniform grid, 
which is generated using the following function:
\begin{equation}
   x_i = \frac{1}{2} \frac{\tanh (\alpha \eta)}{\tanh (\alpha)}, \quad 
   \eta = 2 \frac{i-1}{N-1} - 1,
\end{equation}
where $i$ represents a grid point and $\alpha = 1$. 
The $y$-coordinate is also generated using the same function. 
Grid points with $N = 41$ and 81 are used for this analysis. 
The minimum grid widths in each grid are $\Delta_\mathrm{min} = 0.001H$ and $0.0005H$, respectively. 
The computational region in the $z$-direction is set to the minimum grid width. 
The reference values used for non-dimensionalization are 
$l_\mathrm{ref} = H$, $u_\mathrm{ref} = \alpha/H$, and $T_\mathrm{ref} = T_C$. 
The temperature difference is defined as $\Delta T = T_H-T_C$. 
In this calculation, to compare with existing studies \citep{Davis_1983, Barakos_et_al_1994}, 
the Rayleigh numbers are set to $Ra = 10^3$, $10^4$, $10^5$, and $10^6$. 
The Prandtl number is $Pr = 0.71$. 
The Courant number is $\mathrm{CFL} = \Delta t u_\mathrm{ref}/\Delta_\mathrm{min} = 0.2$.

\begin{figure}[!t]
\begin{minipage}[t]{0.33\hsize}
\begin{center}
\includegraphics[trim=0mm 5mm 0mm 0mm, clip, width=50mm]{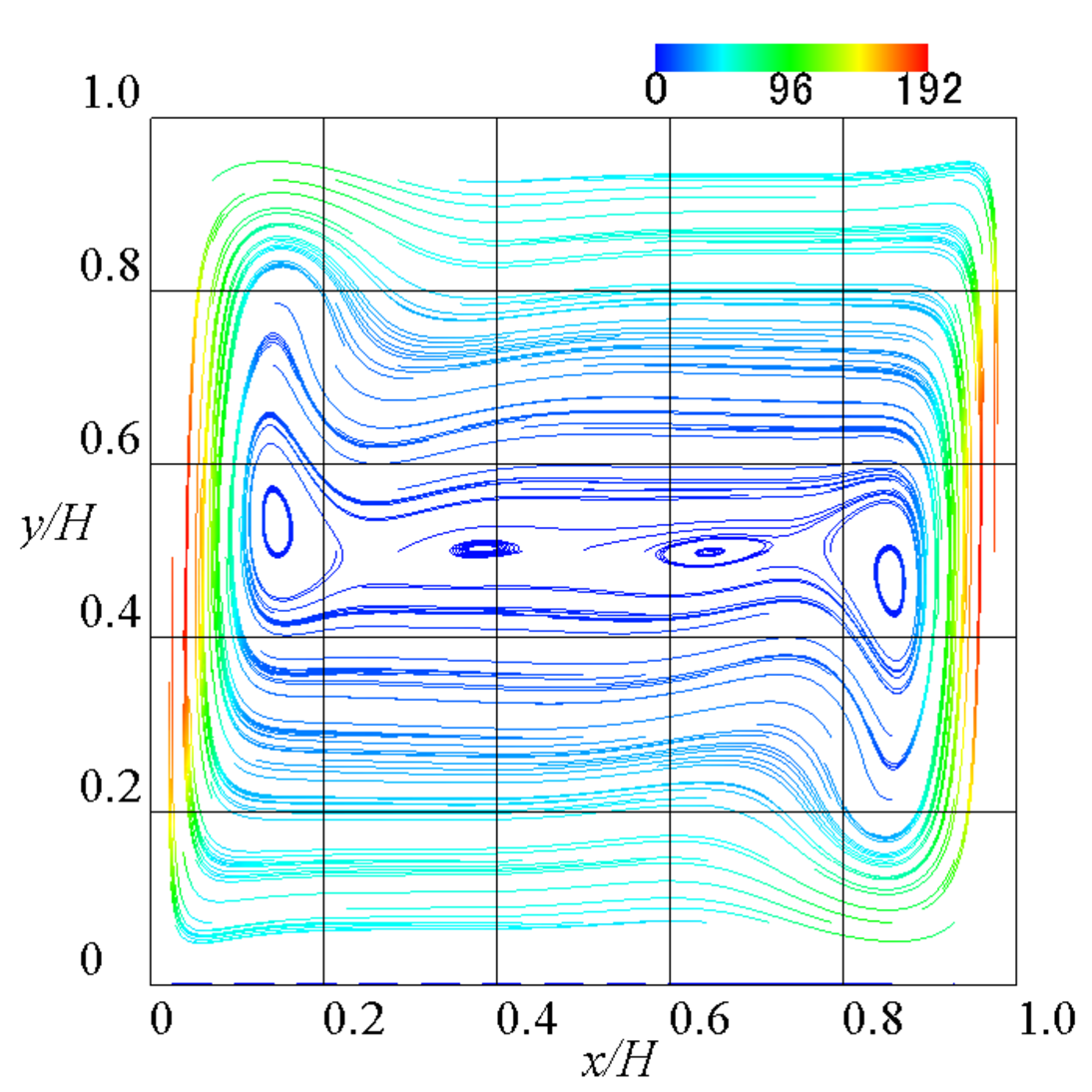} \\
(a) Streamline
\end{center}
\end{minipage}
\begin{minipage}[t]{0.33\hsize}
\begin{center}
\includegraphics[trim=0mm 5mm 0mm 0mm, clip, width=50mm]{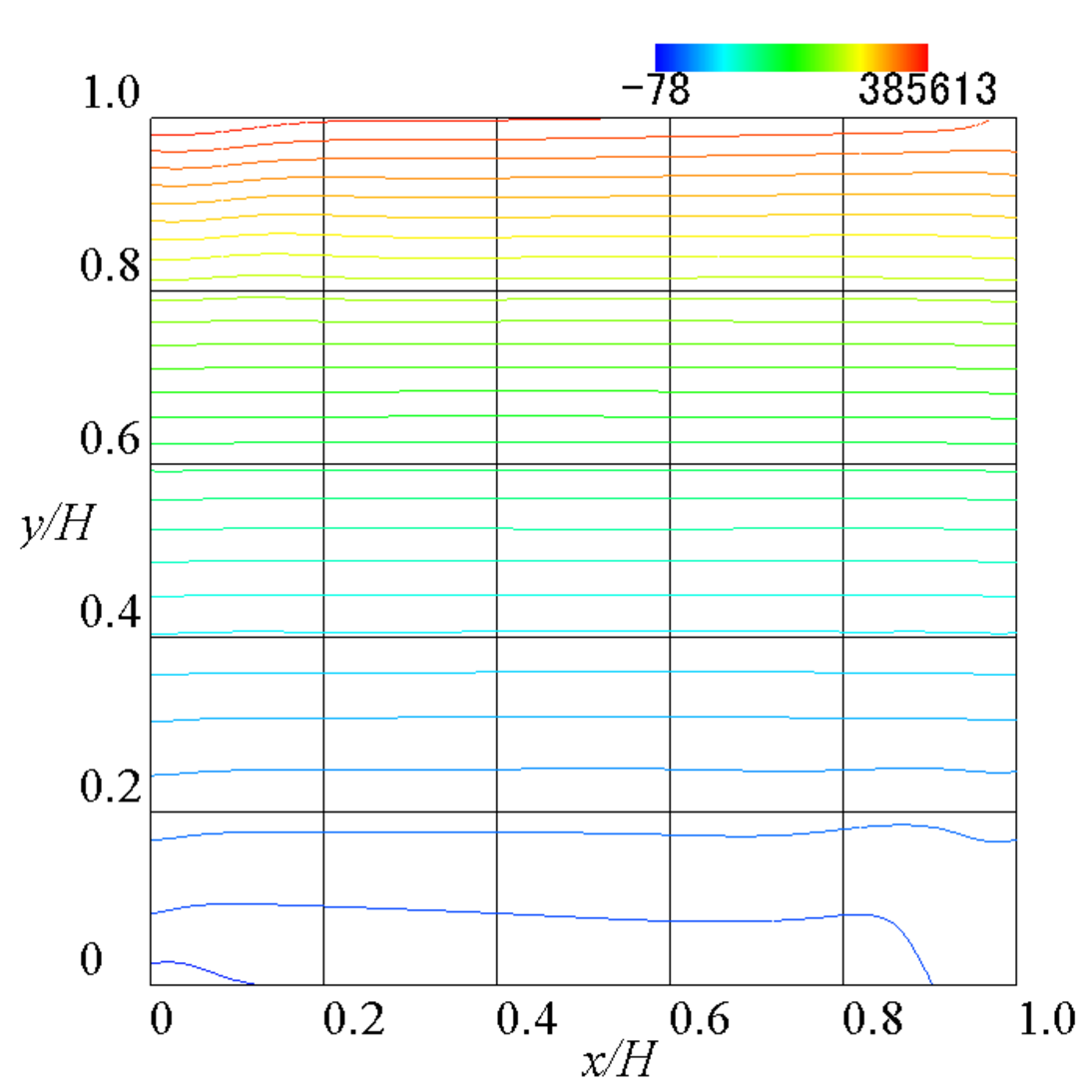} \\
(b) Pressure contour
\end{center}
\end{minipage}
\begin{minipage}[t]{0.33\hsize}
\begin{center}
\includegraphics[trim=0mm 5mm 0mm 0mm, clip, width=50mm]{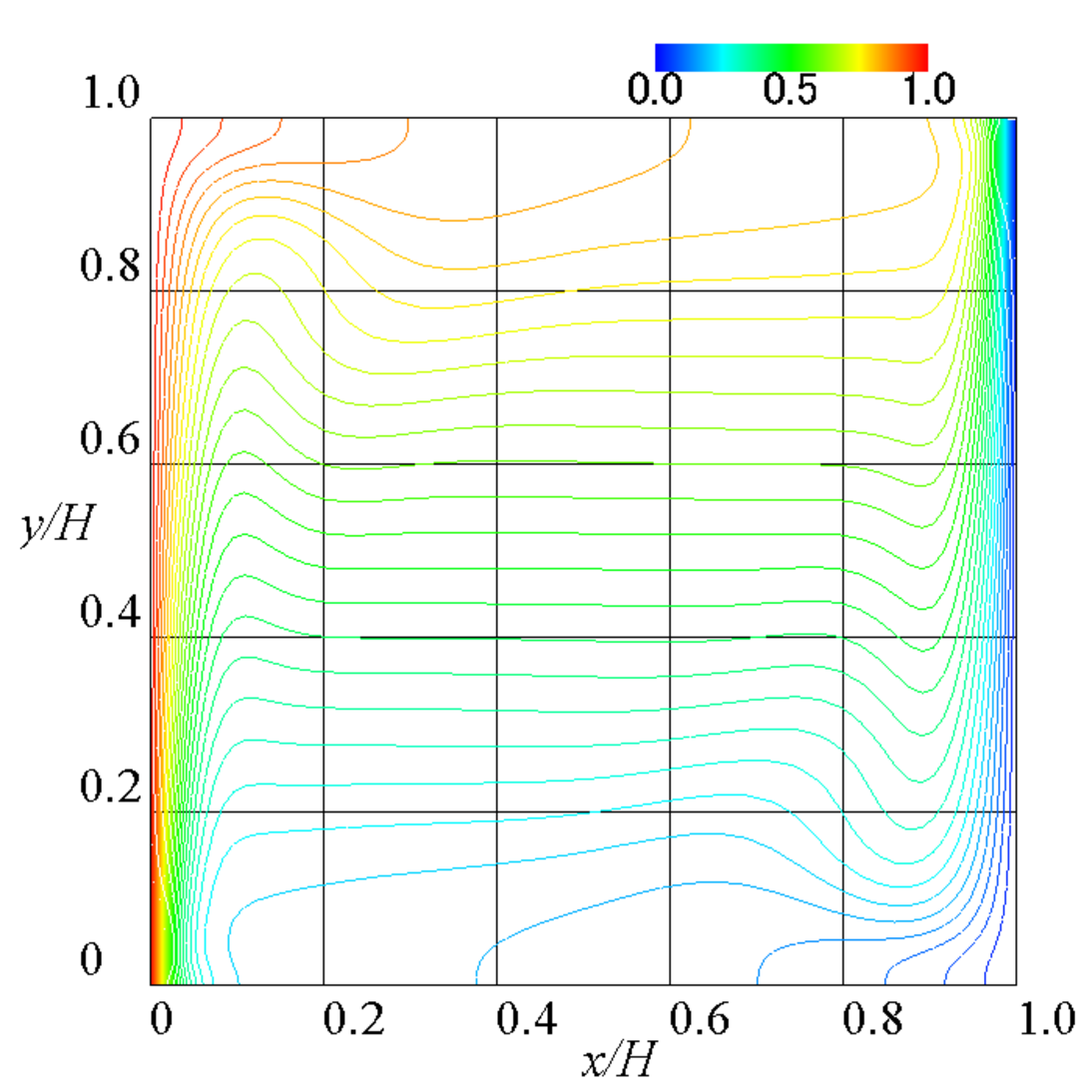} \\
(c) Temperature contour
\end{center}
\end{minipage}
\caption{Flow field at $Ra = 1 \times 10^6$ for a square cavity model}
\label{natuiral_flow}
\end{figure}

Figure \ref{natuiral_flow} shows the streamlines, pressure, 
and temperature distributions at $Ra = 10^6$. 
Fluid heated near the hot wall is pushed up by buoyancy 
and transported to the cold wall. 
On the other hand, the transported high-temperature fluid descends 
while being cooled by the low-temperature wall 
and flows into the high-temperature wall side again. 
A clockwise heat convection is generated by a series of such movements of the fluid. 
The flow and temperature fields obtained using this computational method 
are qualitatively similar to the existing result \citep{Davis_1983, Barakos_et_al_1994}. 
No oscillations are seen in the pressure distribution.

\begin{figure}[!t]
\begin{center}
\includegraphics[trim=0mm 0mm 0mm 0mm, clip, width=70mm]{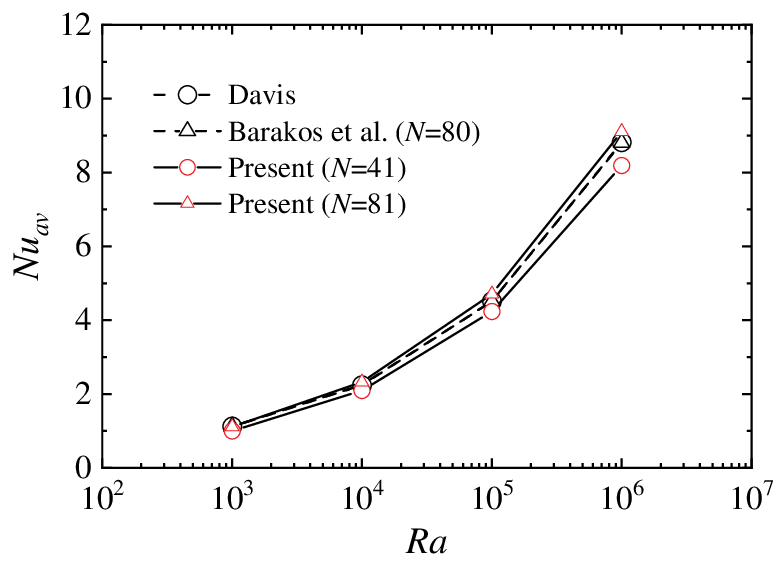}
\end{center}
\vspace*{-1.0\baselineskip}
\caption{Average Nusselt number for a suare cavity model}
\label{natural_Nuav}
\end{figure}

In Fig. \ref{natural_Nuav}, 
the average Nusselt number $Nu_{av}$ on the heating surface is compared 
with the existing values \citep{Davis_1983, Barakos_et_al_1994}. 
In the case of $N = 41$, this calculated value is lower than the previous ones 
and is underestimated compared with the earlier studies. 
For $N = 81$, the present result agrees well with the existing ones. 
It was found from the above results that this computational method can predict 
the heat transfer characteristics at high Rayleigh numbers.

\subsection{Flow in a polar cavity}

\begin{figure}[!t]
\begin{center}
\includegraphics[trim=0mm 15mm 0mm 15mm, clip, width=70mm]{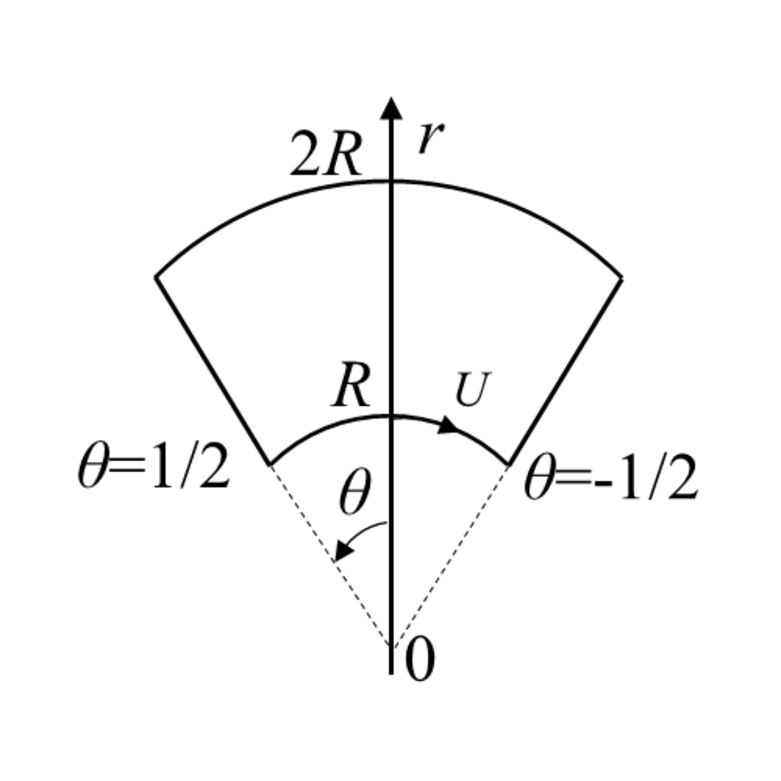}
\end{center}
\vspace*{-0.5\baselineskip}
\caption{Geometry of a polar cavity model}
\label{pcavity}
\end{figure}

Analysis of the flow inside the polar cavity has been carried out 
in existing studies \citep{Zang_et_al_1994, Wu_et_al_1995, Fuchs&Tillmark_1985, Rosenfeld_et_al_1991} 
to verify the numerical method. 
Figure \ref{pcavity} shows the geometry of the polar cavity model. 
The $r$-and $\theta$-axes are in the radial and circumferential directions, respectively, 
and the $z$-axis is perpendicular to the plane of the paper. 
All boundaries are enclosed by walls. 
The internal fluid is driven by the wall moving with uniform velocity $U$ at radius $r = R$. 
Non-slip boundary conditions are given for other wall surfaces. 
The pressure is obtained by second-order accuracy extrapolation. 
Periodic boundary conditions are imposed in the $z$-direction for velocity and pressure. 
The grid used is a $N \times N \times 2$ non-uniform grid, 
which is generated using the following function:
\begin{eqnarray}
   && r_j = R + \frac{1}{2} \frac{\tanh (\alpha \eta_r)}{\tanh (\alpha)}, \quad 
   \eta_r = 2 \frac{j-1}{N-1} - 1, 
   \label{polar_grid_r} \\
   && \theta_i = \frac{1}{2} \frac{\tanh (\alpha \eta_{\theta})}{\tanh (\alpha)}, \quad 
   \eta_{\theta} = 2 \frac{i-1}{N-1} - 1, 
   \label{polar_grid_t} \\
   && x_{i,j} = r_j \sin(\theta_i), \quad 
   y_{i,j} = r_j \cos(\theta_i),
\end{eqnarray}
where $i$ and $j$ represent grid points and $\alpha = 1$. 
Grid points with $N = 41$ and 81 are used for this analysis. 
The minimum grid widths in the radial and circumferential directions are 
$\Delta_\mathrm{min} = 0.01R$ and $0.005R$ for $N = 41$ and 81, respectively. 
The computational domain in the $z$-direction is $0.01R$. 
The reference values used for non-dimensionalization are 
$l_\mathrm{ref} = R$ and $u_\mathrm{ref} = U$. 
In this calculation, the Reynolds number is set to $Re = 350$ 
to compare with existing research \citep{Wu_et_al_1995, Fuchs&Tillmark_1985}. 
The Courant number is $\mathrm{CFL} = \Delta t U/\Delta_\mathrm{min} = 1$.

\begin{figure}[!t]
\begin{minipage}[t]{0.49\hsize}
\begin{center}
\includegraphics[trim=0mm 10mm 0mm 0mm, clip, width=70mm]{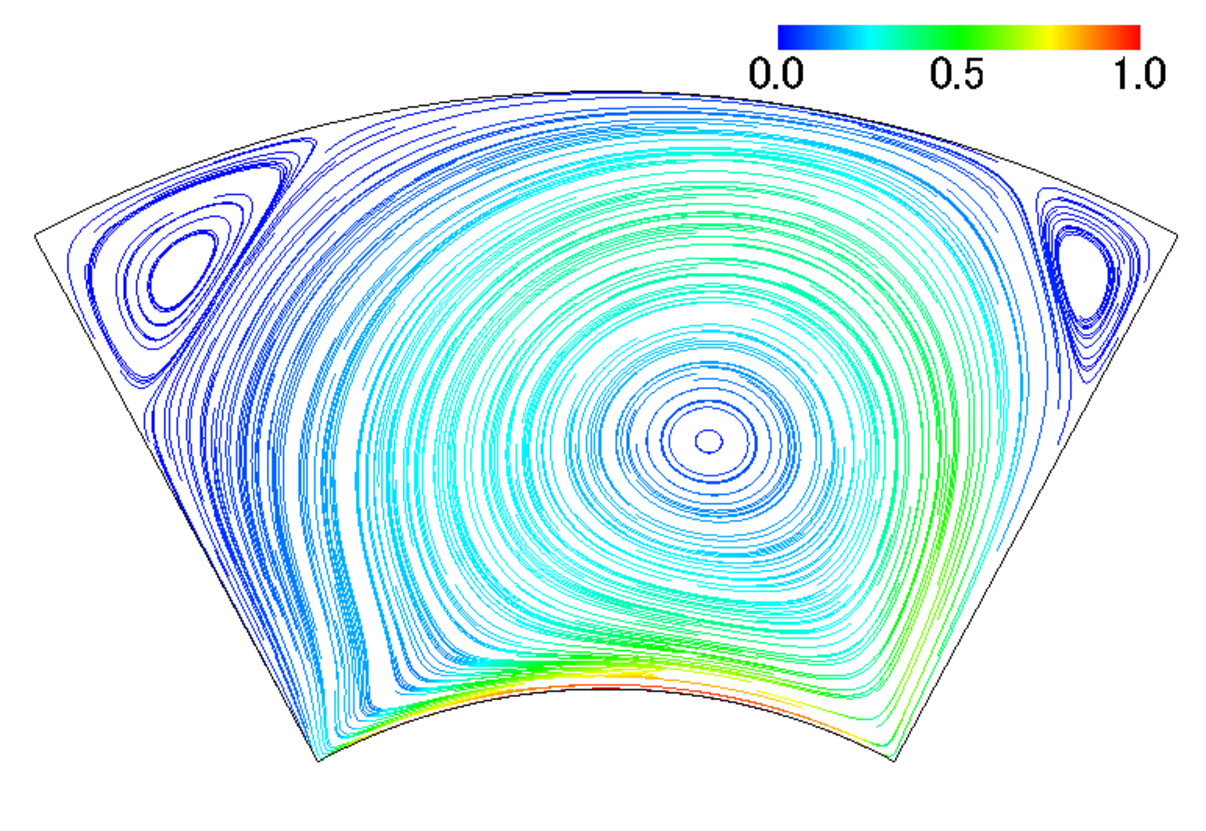} \\
(a) Streamline
\end{center}
\end{minipage}
\begin{minipage}[t]{0.49\hsize}
\begin{center}
\includegraphics[trim=0mm 10mm 0mm 0mm, clip, width=70mm]{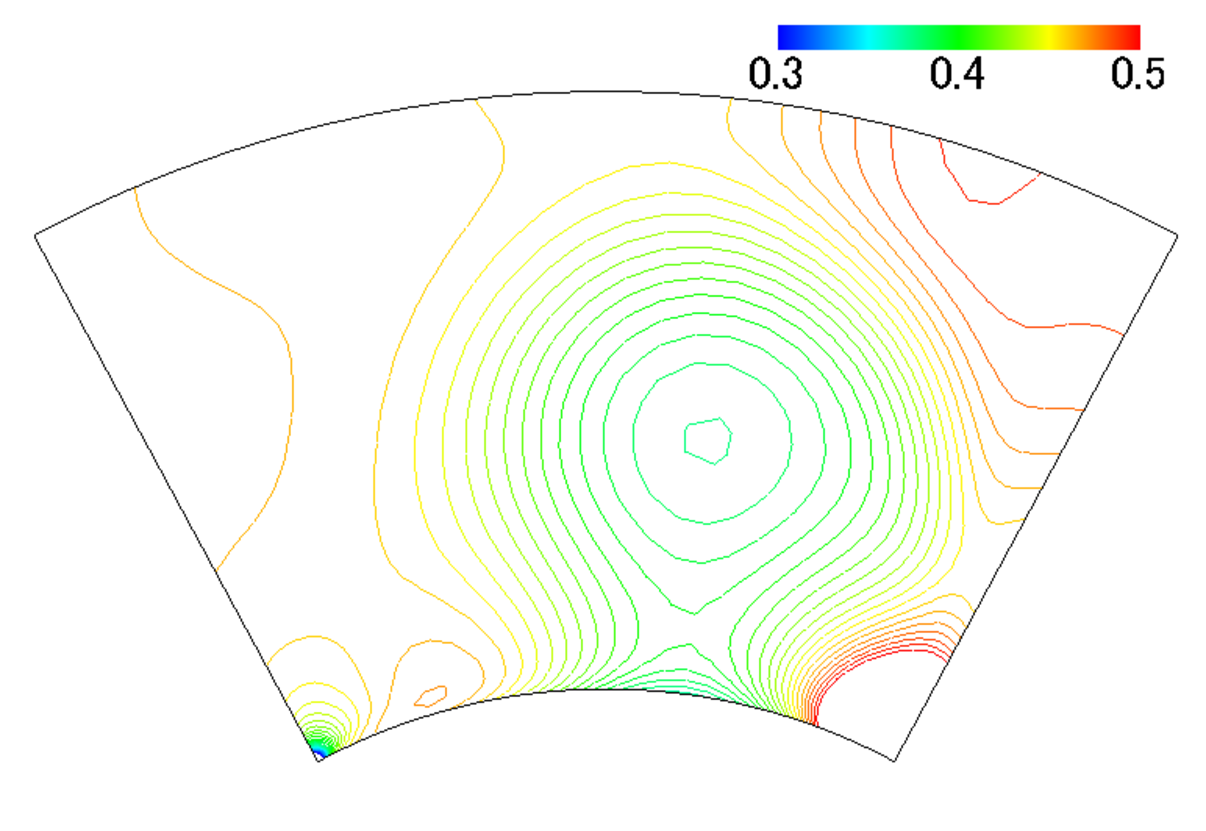} \\
(b) Pressure contour
\end{center}
\end{minipage}
\caption{Flow field at $Re = 350$ for a polar cavity model}
\label{pcavity_flow}
\end{figure}

Streamlines and pressure contours are shown in Fig. \ref{pcavity_flow}. 
There is a large-scale vortex at the center of the flow field 
and secondary vortices at the corners. 
These computational results agree well with the flow visualization experiments of \citet{Fuchs&Tillmark_1985}. 
In addition, no vibration occurs in the pressure distribution. 
In this model, pressure oscillation did not appear 
even without using the Rhie--Chow interpolation.

\begin{figure}[!t]
\begin{minipage}[t]{0.33\hsize}
\begin{center}
\includegraphics[trim=0mm 0mm 0mm 0mm, clip, width=55mm]{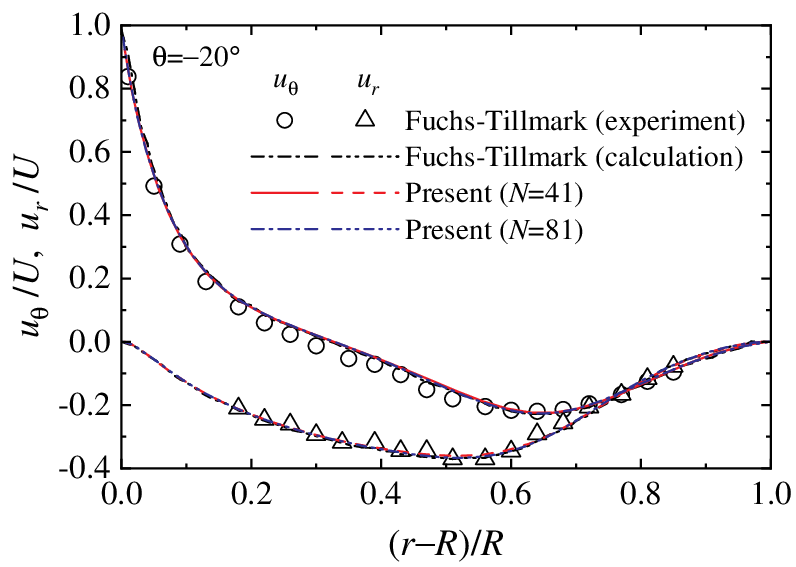} \\
(a) $\theta = -20^{\circ}$
\end{center}
\end{minipage}
\begin{minipage}[t]{0.33\hsize}
\begin{center}
\includegraphics[trim=0mm 0mm 0mm 0mm, clip, width=55mm]{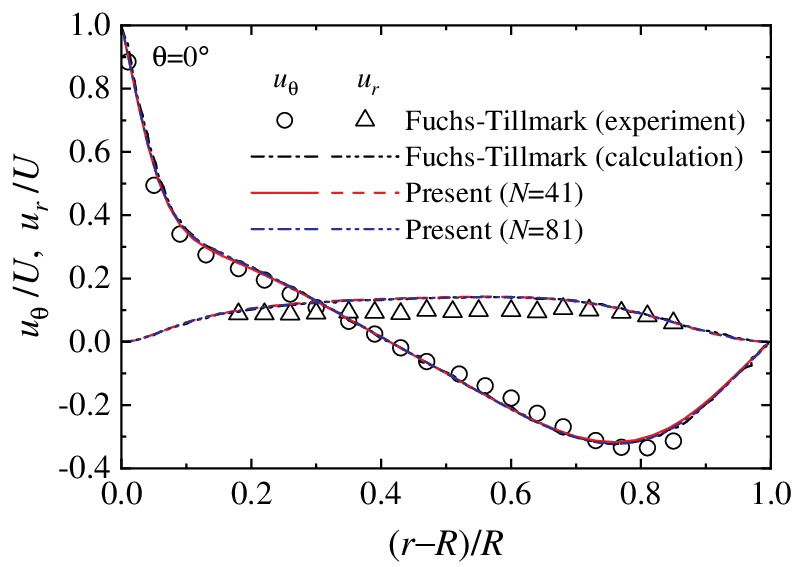} \\
(b) $\theta = 0^{\circ}$
\end{center}
\end{minipage}
\begin{minipage}[t]{0.33\hsize}
\begin{center}
\includegraphics[trim=0mm 0mm 0mm 0mm, clip, width=55mm]{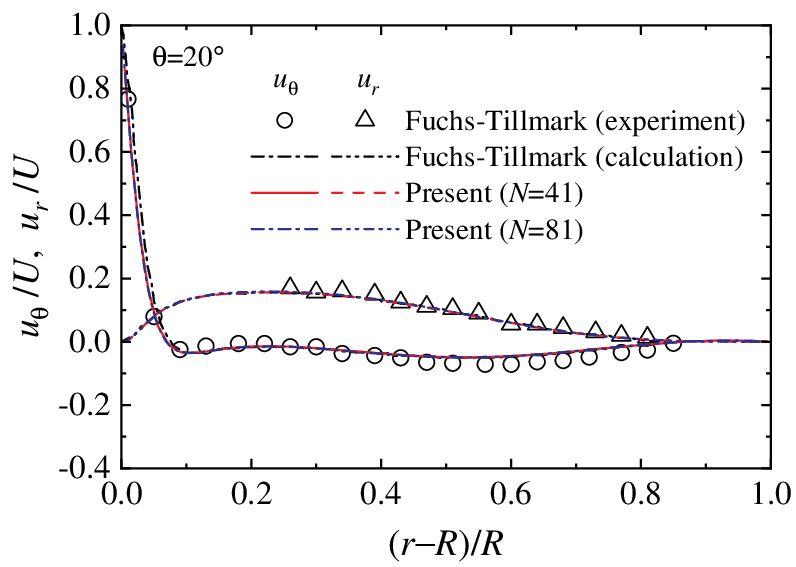} \\
(c) $\theta = 20^{\circ}$
\end{center}
\end{minipage}
\caption{Velocity profiles along radial lines at $Re = 350$ for a polar cavity model}
\label{pcavity_velocity}
\end{figure}

Figure \ref{pcavity_velocity} shows the radial velocity $u_r$ and circumferential velocity $u_{\theta}$ 
at $\theta = -20^{\circ}$, $0^{\circ}$, $20^{\circ}$. 
The experimental and calculated values of \citet{Fuchs&Tillmark_1985} are compared. 
The results obtained using the grid with $N = 41$ and $N = 81$ agree, 
and there is no grid dependency on the calculation results. 
This calculation result is in good agreement with the previous experimental value. 
The existing calculation results were obtained using a grid of $80 \times 80$. 
Although the present calculation results were obtained using about half the grid points 
compared to the previous calculation, 
the distributions for all $ \theta$ agree well with the existing values.

\subsection{Flow in a skewed cavity}

\begin{figure}[!t]
\begin{center}
\includegraphics[trim=0mm 20mm 0mm 20mm, clip, width=70mm]{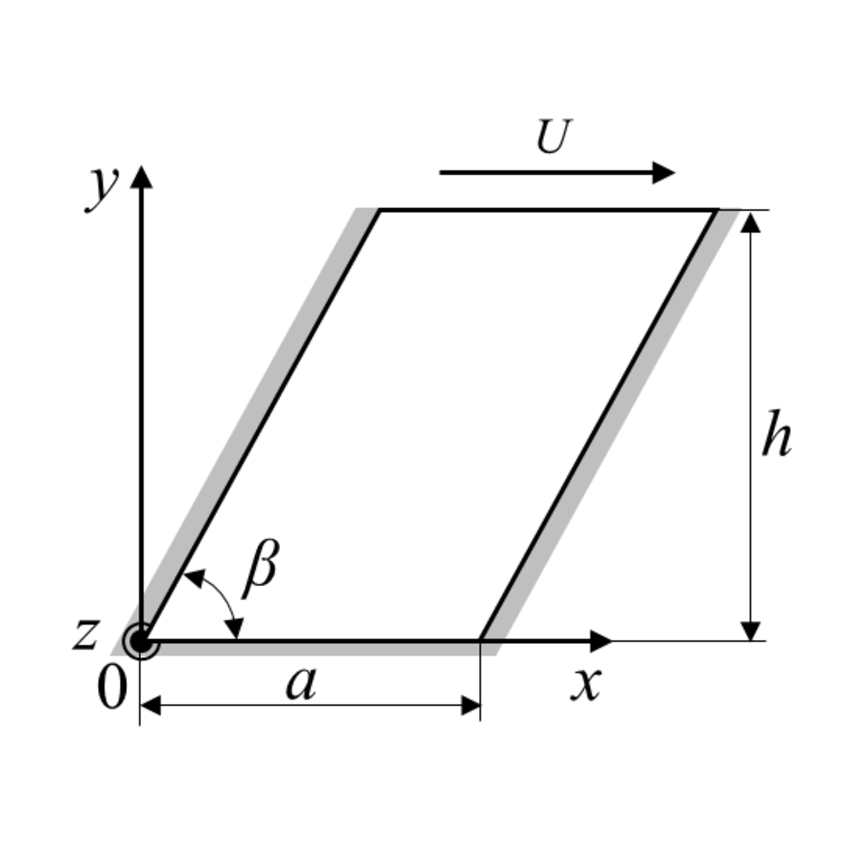}
\end{center}
\vspace*{-0.5\baselineskip}
\caption{Geometry of a skewed cavity model}
\label{scavity}
\end{figure}

Next, to verify the accuracy of this numerical method in non-orthogonal grids, 
we analyze the flow in a skewed cavity, similar to existing research \citep{Peric_1990, Wu_et_al_1995}. 
Figure \ref{scavity} shows the geometry of the skewed cavity model. 
The $x$- and $y$-axes are in the horizontal and vertical directions, respectively, 
and the $z$-axis is perpendicular to the plane of the paper. 
The walls on both sides are inclined at an angle $\beta$, 
and the cavity has a width of $a$ and a height of $h = a \tan \beta$. 
As for the boundary conditions, the upper wall moves with a uniform velocity $U$, 
and non-slip boundary conditions are given for other wall surfaces. 
The pressure is obtained by second-order accuracy extrapolation. 
Periodic boundary conditions are imposed in the $z$-direction for velocity and pressure. 
In this calculation, we analyze the case of $\beta = 30^{\circ}$ and $45^{\circ}$. 
The grid used for the calculation is a non-uniform grid of $61 \times 61 \times 2$ 
and was generated using the same functions as the formulas (\ref{polar_grid_r}) and (\ref{polar_grid_t}). 
The minimum grid width is $\Delta y_\mathrm{min} = 0.0054a$ and $0.0094a$ 
for $\beta = 30^{\circ}$ and $45^{\circ}$, respectively. 
The computational area in the $z$-direction is $0.01a$. 
The number of grid points in the $x$-$y$ cross-section is the same as 
that used in the calculation of \citet{Wu_et_al_1995}. 
The reference values used for non-dimensionalization are 
$l_\mathrm{ref} = a$ and $u_\mathrm{ref} = U$. 
In this calculation, the Reynolds number is set to $Re = 100$ 
to compare with existing research \citep{Peric_1990, Wu_et_al_1995}. 
The time step is $\Delta t/(a/U) = 0.005$, and the Courant number defined as 
$\mathrm{CFL} = \Delta t U/\Delta_\mathrm{min}$ is CFL = 0.92 and 0.53 
for $\beta = 30^{\circ}$ and $45^{\circ}$, respectively.

\begin{figure}[!t]
\begin{minipage}[t]{0.49\hsize}
\begin{center}
\includegraphics[trim=0mm 5mm 0mm 5mm, clip, width=70mm]{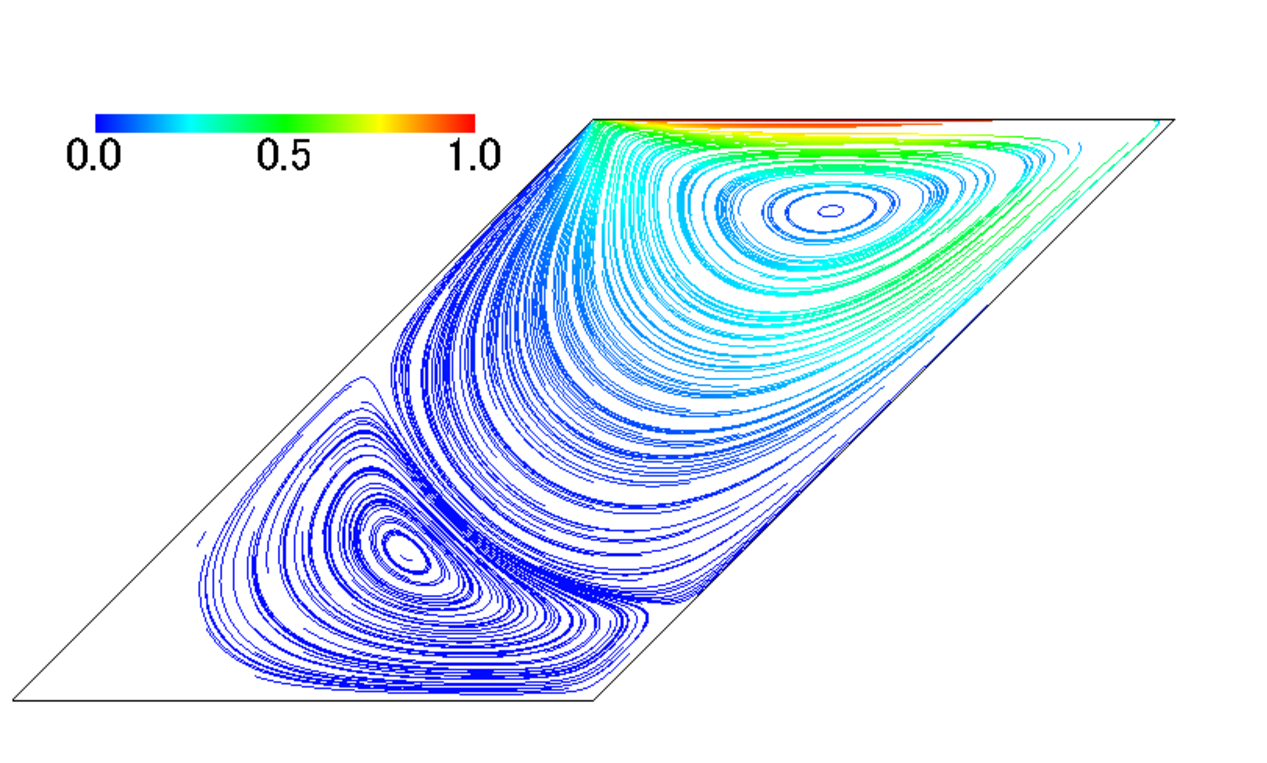} \\
(a) Streamline
\end{center}
\end{minipage}
\begin{minipage}[t]{0.49\hsize}
\begin{center}
\includegraphics[trim=0mm 5mm 0mm 5mm, clip, width=70mm]{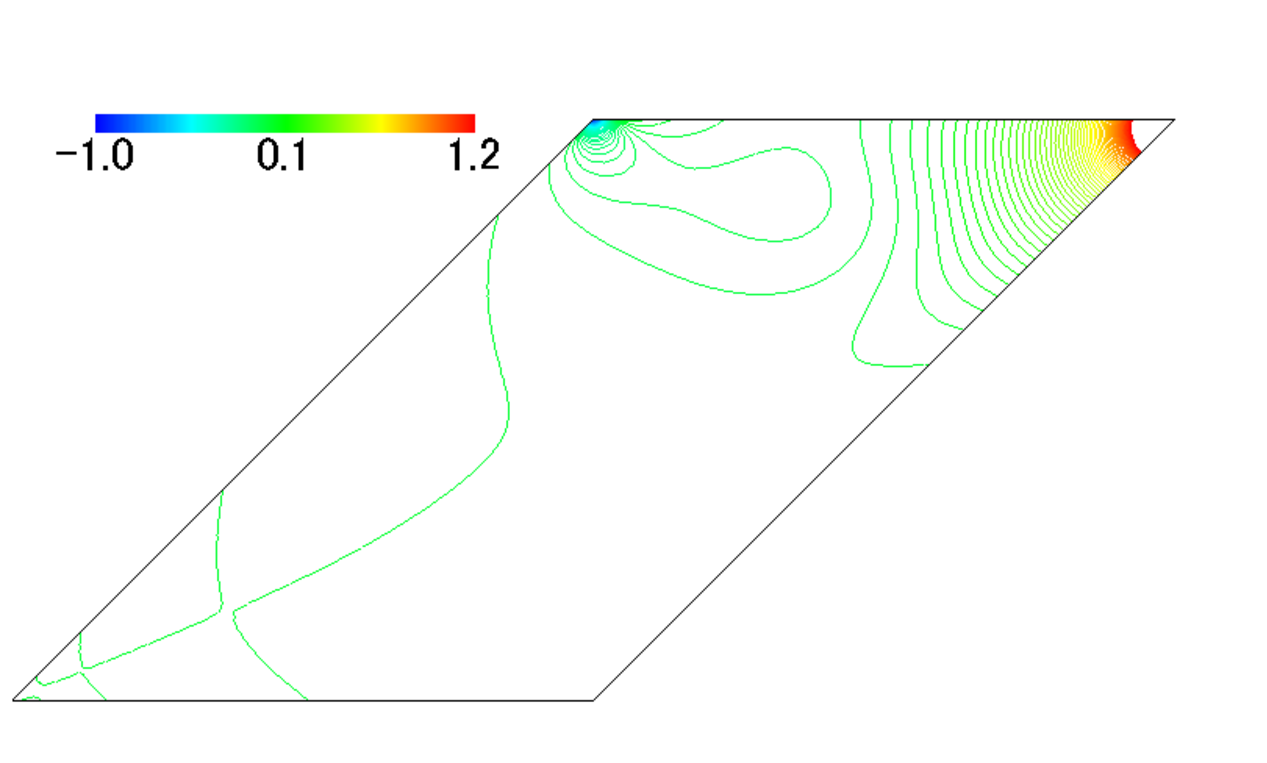} \\
(b) Pressure contour
\end{center}
\end{minipage}
\caption{Flow field at $Re = 100$ and $\beta = 45^{\circ}$ for a skewed cavity model}
\label{scavity_flow_b45}
\end{figure}

The streamlines and pressure contours at $\beta = 45^{\circ}$ are shown in Fig. \ref{scavity_flow_b45}. 
Large-scale and secondary vortices exist at the upper and lower of the flow field, respectively. 
This result qualitatively agrees well with the calculation result of \citet{Peric_1990, Wu_et_al_1995}. 
In addition, no oscillations occur in the pressure distribution. 
In this model, pressure oscillations appeared without the Rhie--Chow interpolation.

\begin{figure}[!t]
\begin{minipage}[t]{0.49\hsize}
\begin{center}
\includegraphics[trim=0mm 0mm 0mm 0mm, clip, width=70mm]{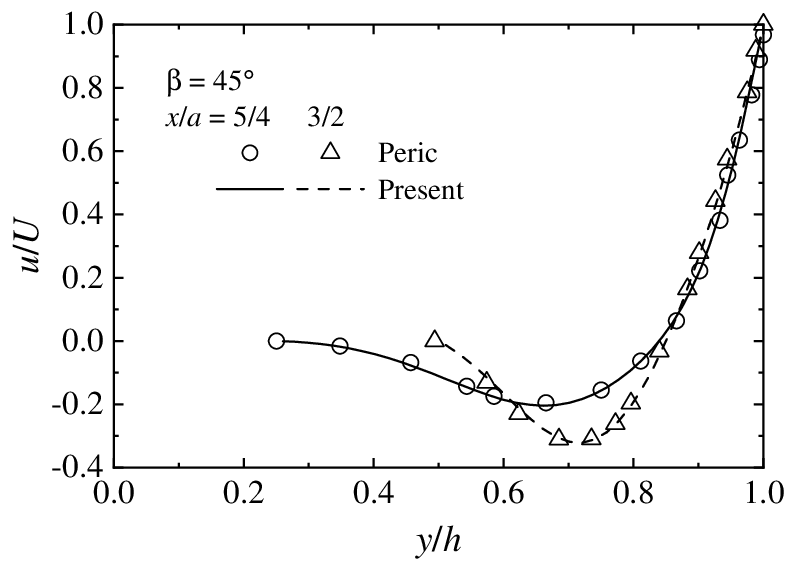} \\
(a) $u/U$
\end{center}
\end{minipage}
\begin{minipage}[t]{0.49\hsize}
\begin{center}
\includegraphics[trim=0mm 0mm 0mm 0mm, clip, width=70mm]{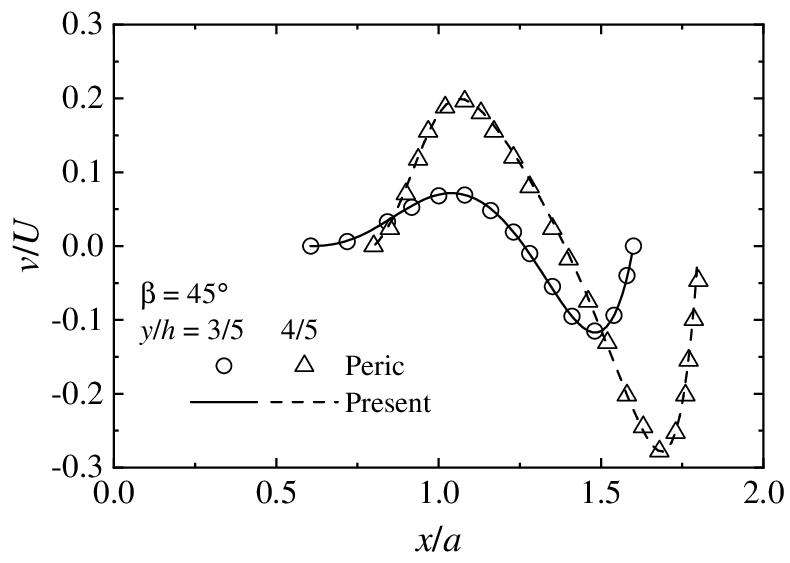} \\
(b) $v/U$
\end{center}
\end{minipage}
\caption{Horizontal and vertical velocity profiles 
at $Re = 100$ and $\beta = 45^{\circ}$ for a skewed cavity model}
\label{scavity_velocity_b45}
\end{figure}

Figure \ref{scavity_velocity_b45} shows the horizontal velocities $u$ at $x/a = 5/4$ and 3/2, 
and the vertical velocities $v$ at $y/h = 3/5$ and 4/5. 
The calculated values of \citet{Peric_1990} are compared. 
The existing calculation results were obtained using a uniform cell of $80\times80$. 
Although the present calculation results used fewer grid points than the existing calculations, 
they agree well with the existing results.

\begin{figure}[!t]
\begin{minipage}[t]{0.49\hsize}
\begin{center}
\includegraphics[trim=0mm 5mm 0mm 50mm, clip, width=70mm]{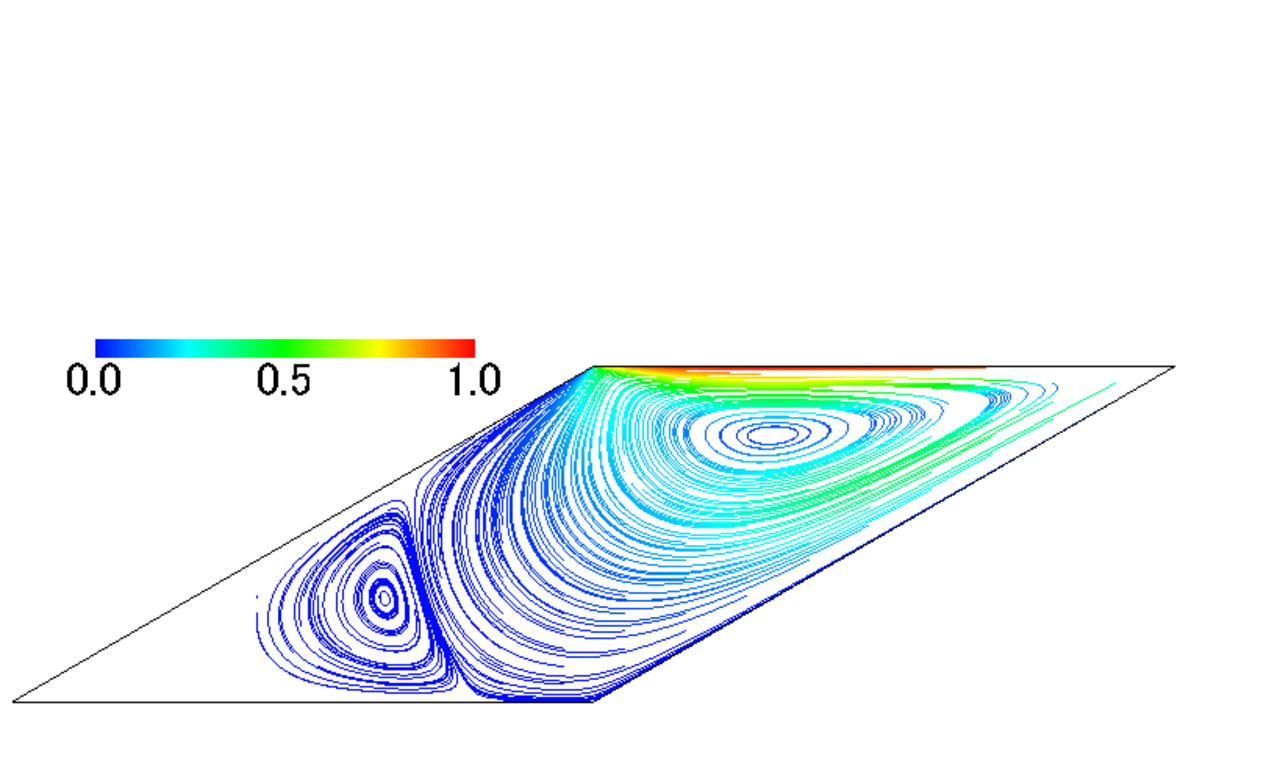} \\
(a) Streamline
\end{center}
\end{minipage}
\begin{minipage}[t]{0.49\hsize}
\begin{center}
\includegraphics[trim=0mm 5mm 0mm 50mm, clip, width=70mm]{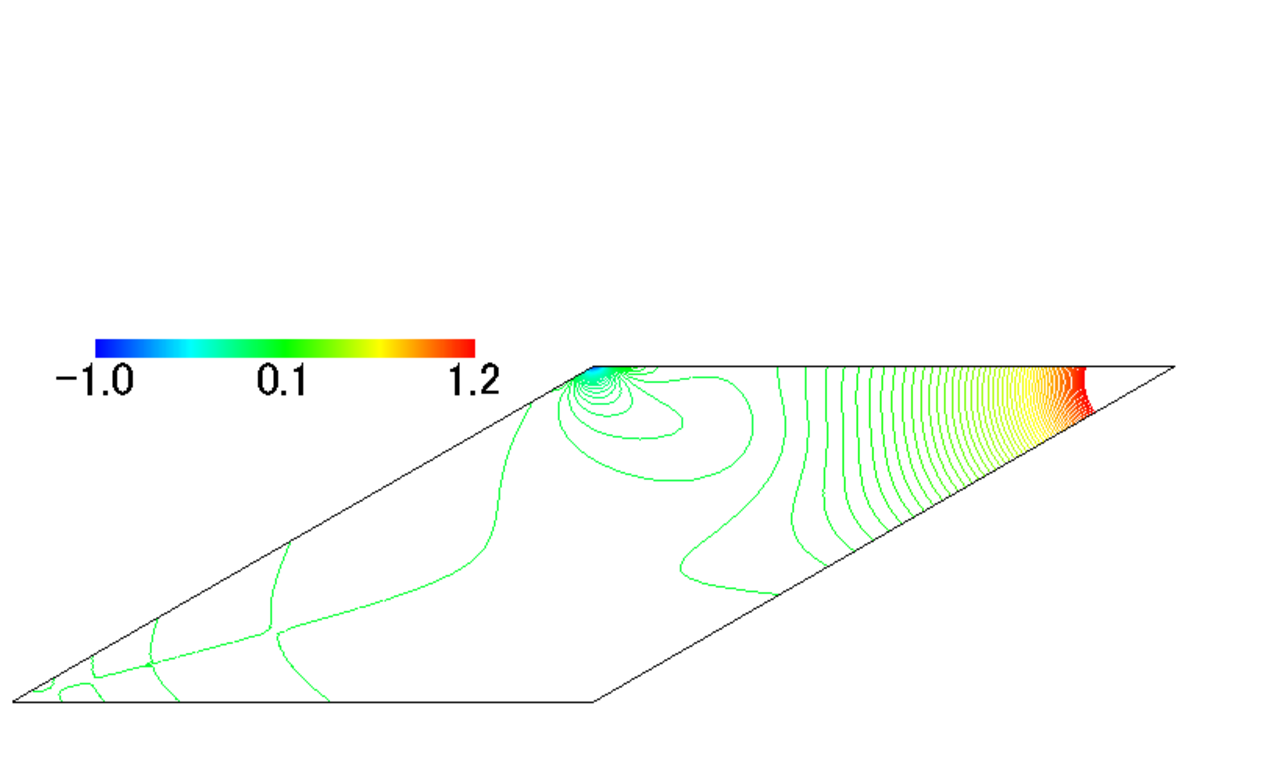} \\
(b) Pressure contour
\end{center}
\end{minipage}
\caption{Flow field at $Re = 100$ and $\beta = 30^{\circ}$ for a skewed cavity model}
\label{scavity_flow_b30}
\end{figure}

The streamlines and pressure contours at $\beta = 30^{\circ}$ are shown in Fig. \ref{scavity_flow_b30}. 
Similar to the result of $\beta = 45^{\circ}$, 
the flow field agrees well with the calculation result of \citet{Peric_1990}. 
It can be seen that the vortices at the top of the cavity reach the bottom wall 
compared with the results for $\beta = 45^{\circ}$. 
In addition, no pressure oscillations occur.

\begin{figure}[!t]
\begin{minipage}[t]{0.49\hsize}
\begin{center}
\includegraphics[trim=0mm 0mm 0mm 0mm, clip, width=70mm]{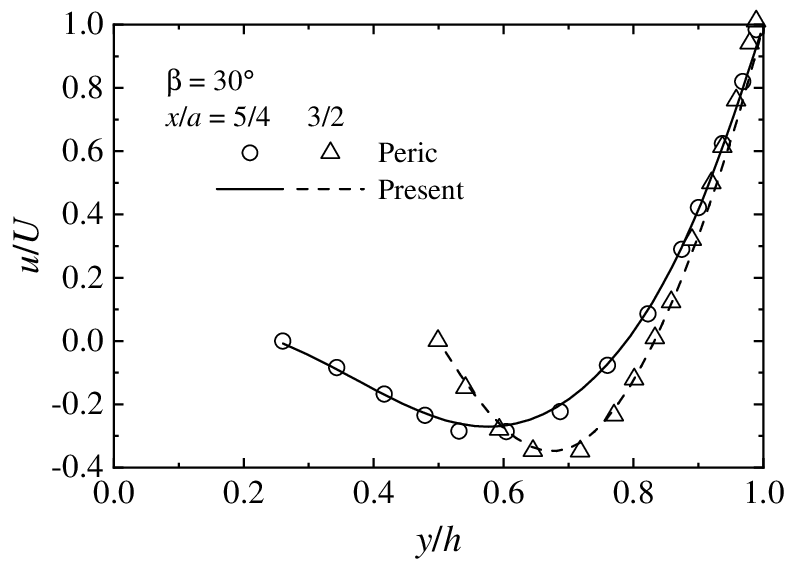} \\
(a) $u/U$
\end{center}
\end{minipage}
\begin{minipage}[t]{0.49\hsize}
\begin{center}
\includegraphics[trim=0mm 0mm 0mm 0mm, clip, width=70mm]{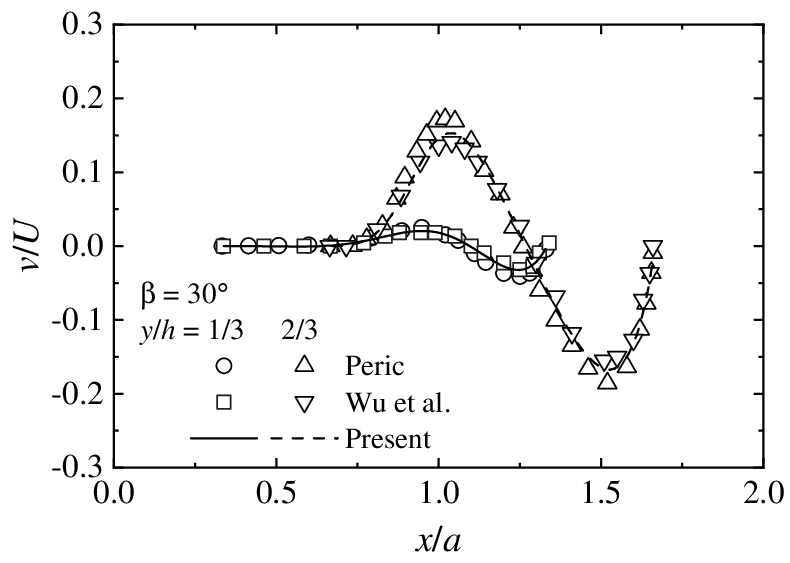} \\
(b) $v/U$
\end{center}
\end{minipage}
\caption{Horizontal and vertical velocity profiles 
at $Re = 100$ and $\beta = 30^{\circ}$ for a skewed cavity model}
\label{scavity_velocity_b30}
\end{figure}

Figure \ref{scavity_velocity_b30} shows the horizontal velocities $u$ at $x/a = 5/4$ and 3/2, 
and the vertical velocities $v$ at $y/h = 1/3$ and 2/3. 
The calculated values of \citet{Peric_1990} and \citet{Wu_et_al_1995} are included for comparison. 
The study of \citet{Wu_et_al_1995} showed the results 
when the cross-derivative term of the Poisson equation was included and ignored. 
Here, the results with the cross-derivative term are compared. 
The present result at $y/h = 2/3$ agrees well with that of \citet{Wu_et_al_1995}, 
although there is a slight difference from the result of \citet{Peric_1990}. 
Overall, it can be said that the calculation results are valid. 
We also performed calculations without omitting the cross-derivative term 
and confirmed that the result matched the present calculated value.

\citet{Wu_et_al_1995} reported that when calculating the Poisson equation without the cross term, 
there was a significant difference from the existing value \citep{Peric_1990}. 
The simultaneous velocity and pressure relaxation method used in the present study 
provides good agreement with existing values even if the cross-derivative term is ignored. 
Furthermore, unlike the analysis of \citet{Peric_1990}, 
it was found that we can perform the calculation stably without any under-relaxation. 
In addition, when calculating without omitting the cross-derivative term at $\beta = 30^{\circ}$, 
the execution time increased by approximately 31%
compared to when the cross term was ignored. 
There is no difference in the decreasing tendency of the residuals of the continuity equations, 
and the residuals decrease to the same level with the same number of iterations. 
The present numerical method makes efficient calculations possible 
even if the cross-derivative term is omitted.

\subsection{Taylor decaying vortex}

For high Reynolds number flows with the decaying of kinetic energy, 
the accuracy of the present numerical method is verified 
by comparing the calculation result with the exact solution. 
A Taylor decaying vortex analysis is performed to verify the accuracy and convergence. 
The solution to the Taylor decaying vortex problem \citep{Taylor_1923} is given as
\begin{eqnarray}
   u &=& - \cos(k x) \sin(k y) e^{-\frac{2 k^2}{Re}t}, \\
   v &=&   \sin(k x) \cos(k y) e^{-\frac{2 k^2}{Re}t}, \\
   p &=& - \frac{1}{4} \left[ \cos(2 k x) + \cos(2 k y) \right] e^{-\frac{4 k^2}{Re}t}
\end{eqnarray}
where $k = 2\pi$. These equations are non-dimensionalized 
by the maximum velocity $U$ and the wavelength $L$ of the periodic vortex. 

The calculation area is $L \times L$, and 
the computational region in the $z$-direction is the grid spacing. 
The exact solution is given as the initial condition, 
and the periodic boundary is set as the boundary condition. 
A uniform grid with $N \times N \times 2$ is used. 
$N$ is the number of grid points in the $x$- and $y$-directions. 
$N$ is changed to $N = 11$, 21, 41, and 81, 
and the convergence of the calculation results 
with respect to the number of grid points is investigated. 
The reference values used in this calculation are 
$l_\mathrm{ref} = L$ and $u_\mathrm{ref} = U$. 
The Reynolds number is changed to $Re = 10^2$, $10^3$, and $5\times10^3$. 
The Courant number is defined as $mathrm{CFL} = \Delta t U/\Delta x$ 
using the maximum velocity $U$ and grid spacing $\Delta x$. 
For inviscid analysis, we use a grid of $N = 41$ 
and set the time step at which the Courant number becomes $\mathrm{CFL} = 0.5$. 
In viscous analysis, the Courant number is $\mathrm{CFL} = 0.4$ for each grid.

\begin{figure}[!t]
\begin{minipage}[t]{0.49\hsize}
\begin{center}
\includegraphics[trim=0mm 0mm 0mm 0mm, clip, width=70mm]{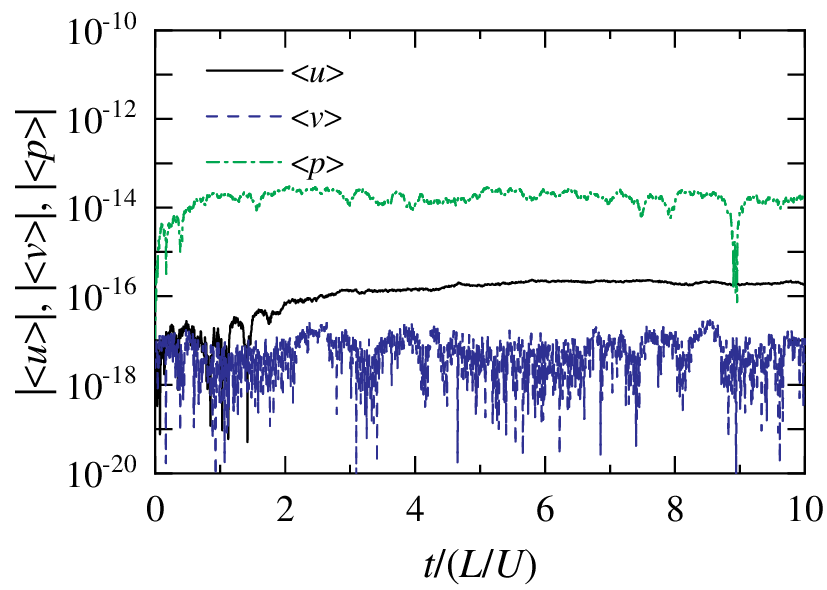} \\
(a) Velocity
\end{center}
\end{minipage}
\begin{minipage}[t]{0.49\hsize}
\begin{center}
\includegraphics[trim=0mm 0mm 0mm 0mm, clip, width=70mm]{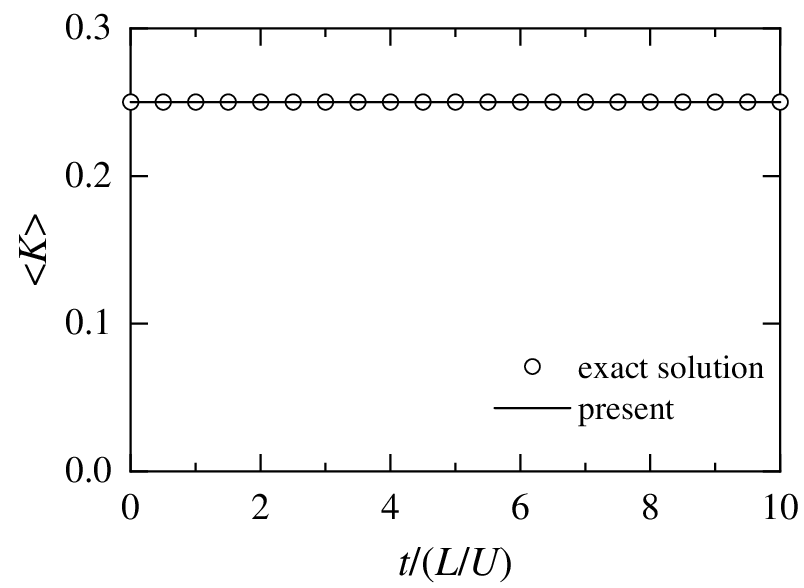} \\
(b) Kinetic energy
\end{center}
\end{minipage}
\caption{Total amounts of velocity and kinetic energy at $Re = \infty$ 
using the present method for a decaying vortex}
\label{decay2d_inviscid_mrc_sum_u&K}
\end{figure}

\begin{figure}[!t]
\begin{minipage}[t]{0.49\hsize}
\begin{center}
\includegraphics[trim=0mm 0mm 0mm 0mm, clip, width=70mm]{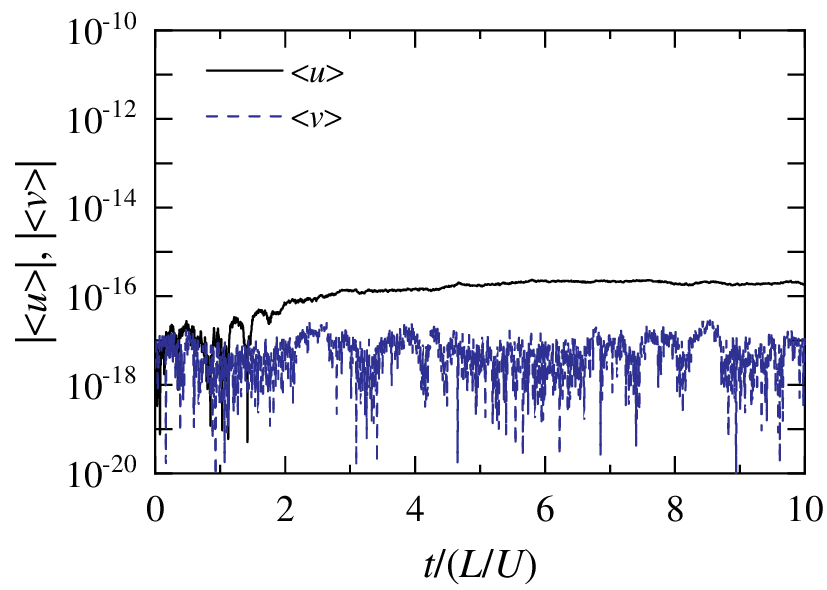} \\
(a) Velocity
\end{center}
\end{minipage}
\begin{minipage}[t]{0.49\hsize}
\begin{center}
\includegraphics[trim=0mm 0mm 0mm 0mm, clip, width=70mm]{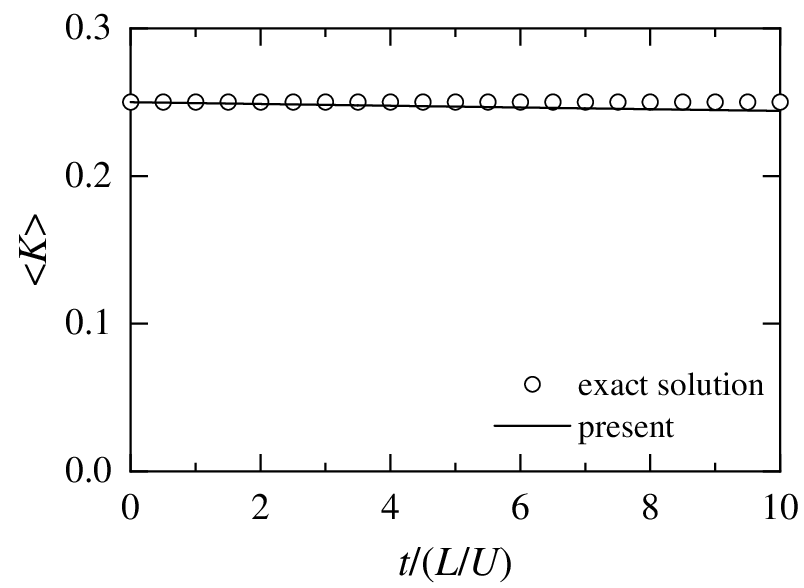} \\
(b) Kinetic energy
\end{center}
\end{minipage}
\caption{Total amounts of velocity and kinetic energy at $Re = \infty$ 
using the Rhie--Chow scheme for a decaying vortex}
\label{decay2d_inviscid_rc_sum_u&K}
\end{figure}

As this computational model is a periodic flow, 
the total amounts of momentum and kinetic energy are conserved for $Re = \infty$. 
Similar to existing research\citep{Yanaoka_2023}, we investigate the conservation properties 
for momentum and kinetic energy. 
Figure \ref{decay2d_inviscid_mrc_sum_u&K}(a) shows the total amount, 
$\langle u \rangle$ and $\langle v \rangle$, 
of velocity obtained using this numerical method. 
The total amount was determined by volume integration within the calculation domain. 
Each total amount is zero, as can be seen from the volume integral of the exact solution. 
All the total amounts remain at low levels, 
indicating excellent conservation of the velocity. 
The total amount, $\langle K \rangle$, of kinetic energy is shown in Fig. \ref{decay2d_inviscid_mrc_sum_u&K}(b). 
This calculation result agrees well with the exact solution, 
and it can be seen that the energy is conserved. 
Figure \ref{decay2d_inviscid_rc_sum_u&K} shows the results obtained by the Rhie--Chow scheme. 
Although the velocity is conserved, 
the total amount of kinetic energy deviates from its initial value over time. 
If $Re = \infty$, the initial value must be maintained. 
In the Rhie-Chow scheme, the kinetic energy contains a first-order accuracy error to time, 
which degrades the energy conservation property.

\begin{figure}[!t]
\begin{minipage}[t]{0.49\hsize}
\begin{center}
\includegraphics[trim=0mm 0mm 0mm 0mm, clip, width=70mm]{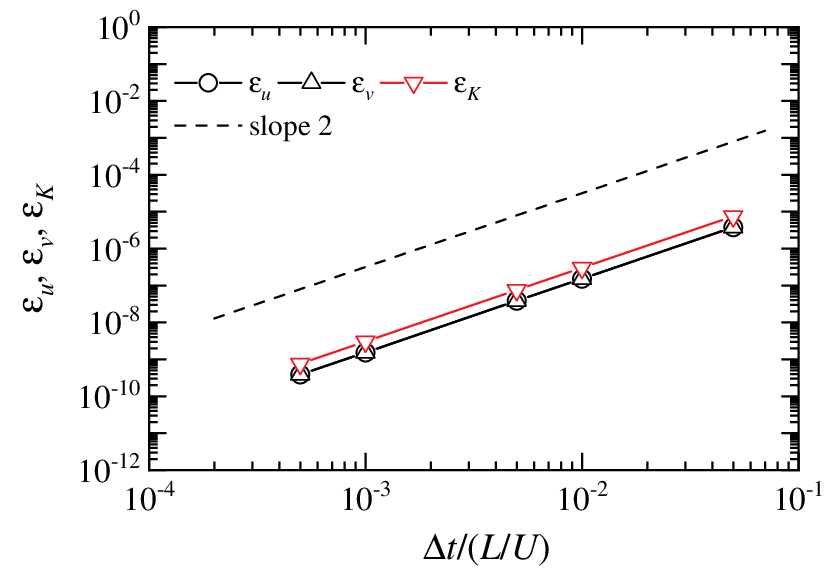} \\
(a) Present
\end{center}
\end{minipage}
\begin{minipage}[t]{0.49\hsize}
\begin{center}
\includegraphics[trim=0mm 0mm 0mm 0mm, clip, width=70mm]{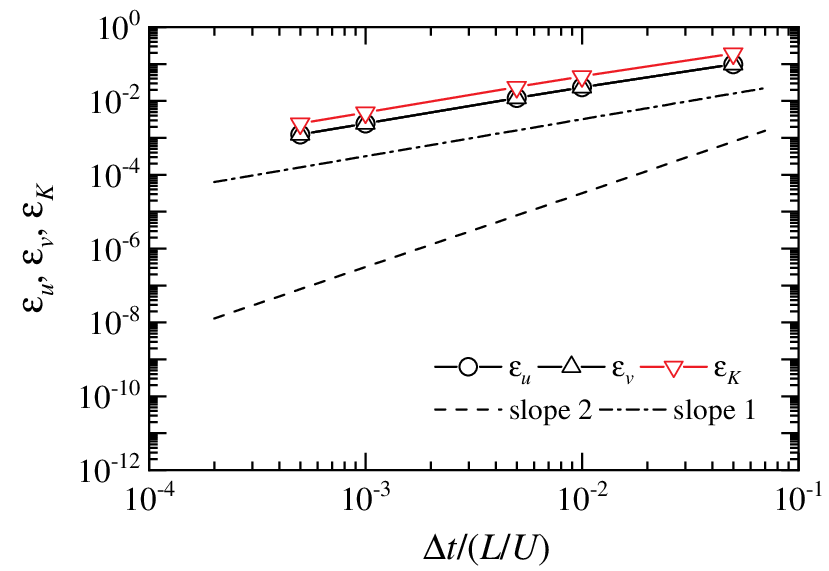} \\
(b) Rhie--Chow scheme
\end{center}
\end{minipage}
\caption{Errors of velocity and kinetic energy at $Re = \infty$ for a decaying vortex}
\label{decay2d_inviscid_error}
\end{figure}

The maximum velocity errors, $\varepsilon_u$ and $\varepsilon_v$, 
and the relative error, $\varepsilon_K$, of kinetic energy are shown in Fig. \ref{decay2d_inviscid_error}. 
The relative error is defined as the relative difference from the initial value. 
When using this numerical method, 
the slope of the errors for velocity and kinetic energy is 2, 
and the error converges to second-order accuracy as the time step $\Delta t$ decreases. 
On the other hand, the Rhie--Chow interpolation method does not show second-order convergence, 
and it is found that errors in first-order accuracy are included 
in velocity and kinetic energy.

\begin{figure}[!t]
\begin{minipage}[t]{0.33\hsize}
\begin{center}
\includegraphics[trim=0mm 0mm 0mm 0mm, clip, width=55mm]{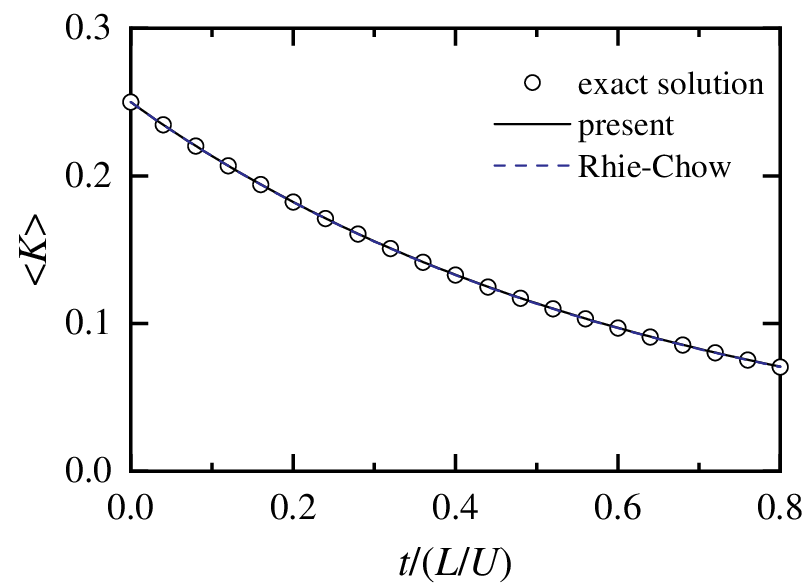} \\
(a) $Re = 100$
\end{center}
\end{minipage}
\begin{minipage}[t]{0.33\hsize}
\begin{center}
\includegraphics[trim=0mm 0mm 0mm 0mm, clip, width=55mm]{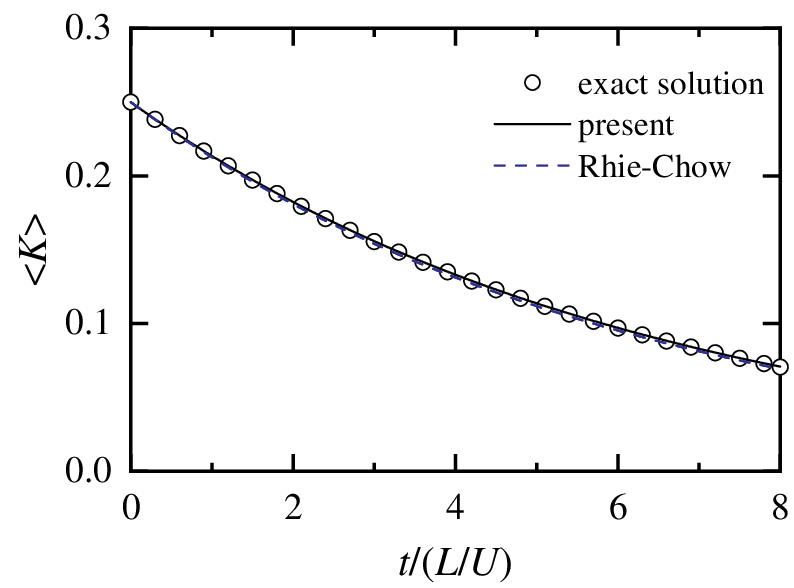} \\
(b) $Re = 1000$
\end{center}
\end{minipage}
\begin{minipage}[t]{0.33\hsize}
\begin{center}
\includegraphics[trim=0mm 0mm 0mm 0mm, clip, width=55mm]{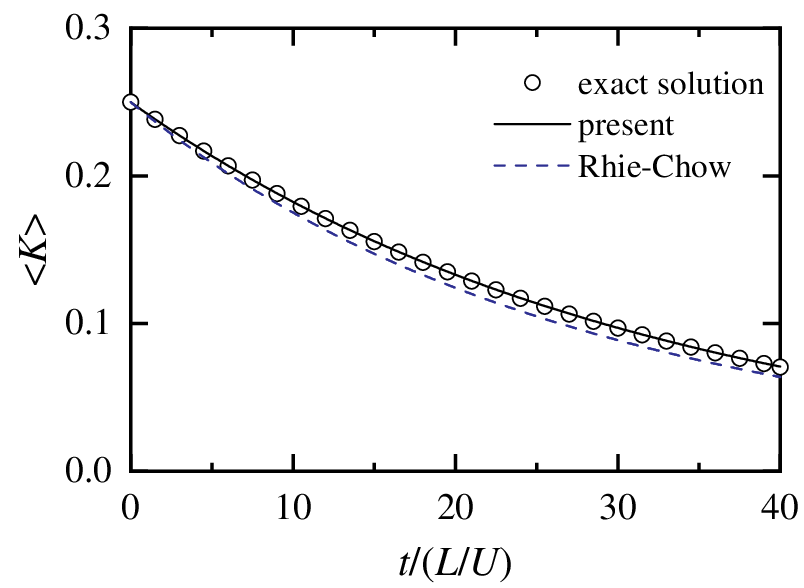} \\
(c) $Re = 5000$
\end{center}
\end{minipage}
\caption{Kinetic energies at $Re=100$, 1000, and 5000 for a decaying vortex: $N = 41$}
\label{decay2d_K_Re}
\end{figure}

Next, we demonstrate the results of the viscous analysis. 
The total amount of kinetic energy at $Re = 100-5000$ is shown in Fig.\ref{decay2d_K_Re}. 
The number of grid points used is $N = 41$. 
For $Re = 100$ and 1000, the results in both interpolation methods agree with the exact solution. 
In a flow dominated by viscosity, 
the influence of the first-order accuracy error contained in the kinetic energy 
does not appear in the time variation of the kinetic energy. 
At $Re = 5000$, there is a difference between the result by Rhie--Chow's method and the exact solution. 
At a high Reynolds number, the difference between the interpolation methods appears.

\begin{figure}[!t]
\begin{minipage}[t]{0.49\hsize}
\begin{center}
\includegraphics[trim=0mm 0mm 0mm 0mm, clip, width=70mm]{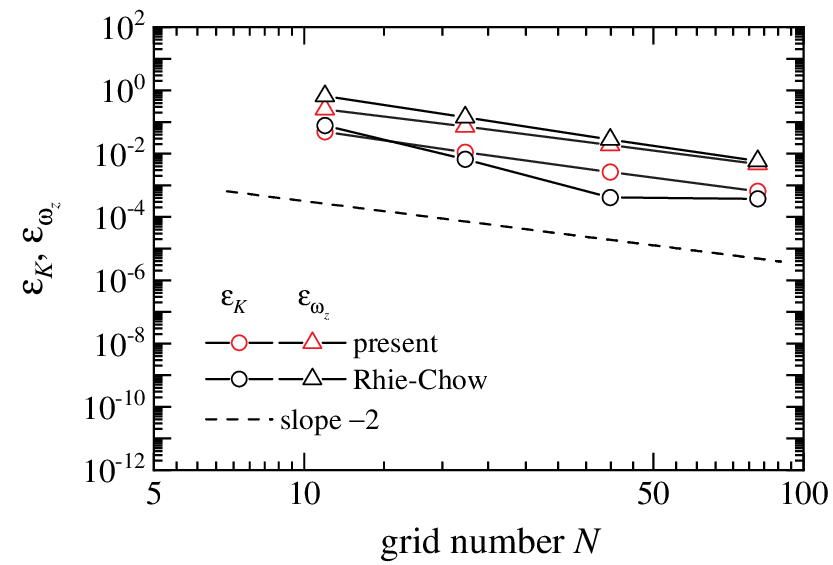} \\
(a) $Re = 100$
\end{center}
\end{minipage}
\begin{minipage}[t]{0.49\hsize}
\begin{center}
\includegraphics[trim=0mm 0mm 0mm 0mm, clip, width=70mm]{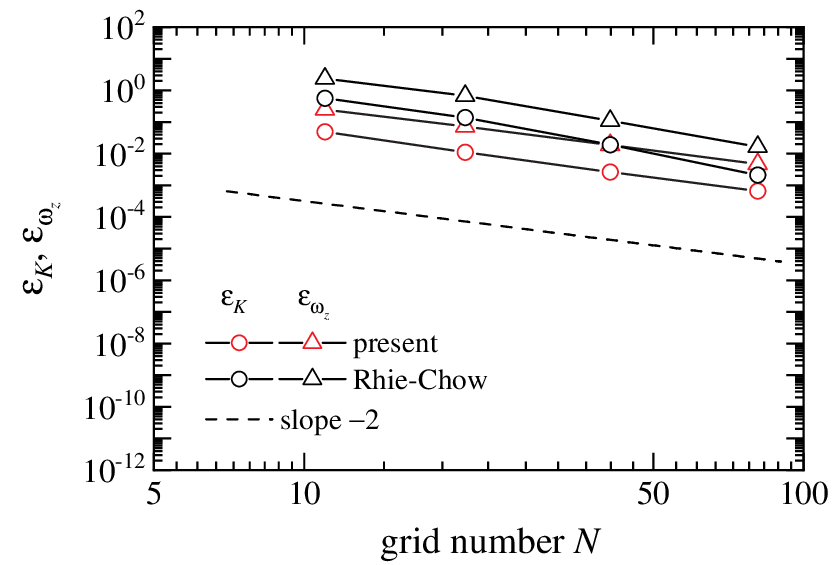} \\
(b) $Re = 1000$
\end{center}
\end{minipage}
\caption{Variations of the relative error of kinetic energy and maximum error of vorticity 
with a number of grids using different interpolations at $Re = 1000$ for a decaying vortex}
\label{decay2d_K_grid}
\end{figure}

Figure \ref{decay2d_K_grid} shows the relative error, 
$\varepsilon_{K} = |(\langle K \rangle-\langle K \rangle_e)/\langle K \rangle_e|$, 
of the kinetic energy. 
Here, the subscript $e$ represents the exact solution. 
The maximum error, $\varepsilon_{\omega_z}$, 
of vorticity in the $z$-direction is also shown. 
When the Reynolds number is low, 
there is almost no difference between the interpolation methods. 
When the Reynolds number increases, 
the error of the present method is lower than that of the Rhie--Chow interpolation method. 
As the number of grid points increases, the error decreases with a slope of $-2$, 
indicating second-order convergence.

As can be seen by substituting the exact solution into the Navier--Stokes equation (\ref{navier-stokes}), 
the time derivative term is canceled by the viscous term. 
Therefore, we could not confirm the effect of time increments on calculation accuracy 
in the viscosity analysis.

\subsection{Periodic three-dimensional inviscid flow}

In the Taylor decaying vortex model, when $Re = \infty$, 
the exact solution does not change over time, 
and the initial value is maintained. 
In this subsection, we analyze a three-dimensional inviscid flow 
in which the velocity and pressure change over time from the initial conditions.

As an initial condition for the three-dimensional flow field, 
the vector potential $\bm{\Psi}$ is given as
\begin{equation}
   \Psi_x = \frac{1}{k} \sin(k y), \quad 
   \Psi_y = \frac{1}{k} \sin(k z), \quad 
   \Psi_z = \frac{1}{k} \sin(k x),
   \label{stream_vector_potential}
\end{equation}
where $k = 2 \pi$. 
Using the relationship $\bm{u} = \nabla \times \bm{\Psi}$, 
the velocity can be obtained from Eq. (\ref{stream_vector_potential}) as follows:
\begin{equation}
   u = - \cos(k z), \quad 
   v = - \cos(k x), \quad 
   w = - \cos(k y),
   \label{velocity}
\end{equation}
The expression (\ref{velocity}) automatically satisfies the divergence-free condition 
$\nabla \cdot \bm{u} = 0$. 
Equation (\ref{velocity}) is made dimensionless using the maximum velocity value $U$ 
as a reference value.

\begin{figure}[!t]
\begin{center}
\includegraphics[trim=0mm 0mm 0mm 0mm, clip, width=70mm]{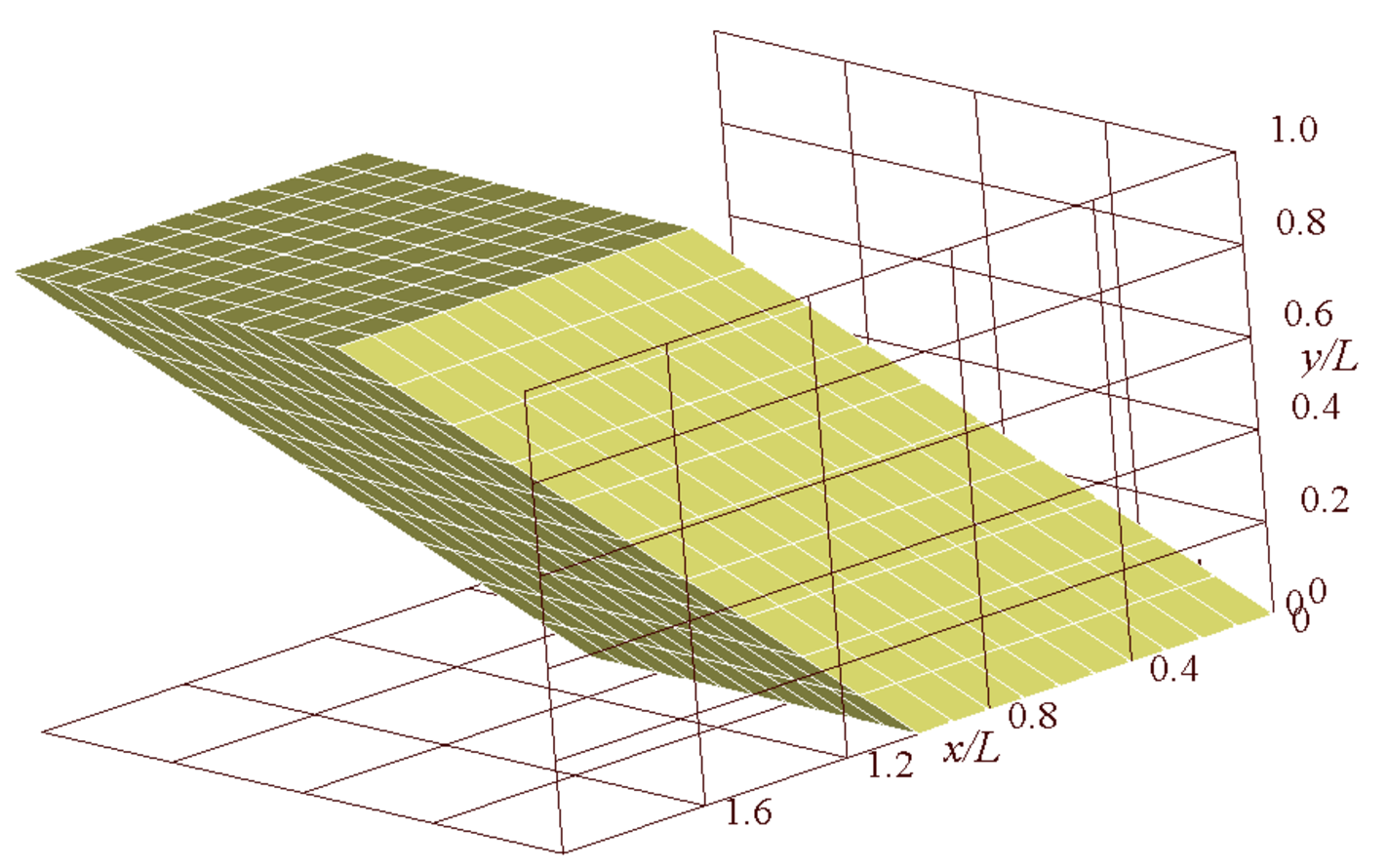}
\end{center}
\vspace*{-0.5\baselineskip}
\caption{Skewed grid for a periodic inviscid flow}
\label{inviscid_grid}
\end{figure}

The computational domain is a cube with a side of $L$. 
As for the initial condition, we give Eq. (\ref{stream_vector_potential}). 
Periodic boundary conditions are applied at all boundaries. 
This calculation uses a uniform grid of $N \times N \times N$. 
$N$ is the number of grid points in each direction, 
and it was fixed at $N = 11$. 
To investigate the effect of grid non-orthogonality on calculation accuracy, 
we use the non-orthogonal grid shown in Fig. \ref{inviscid_grid}. 
The computational domain is tilted at an angle $\beta_1 = 45^{\circ}$ in the $x$-direction 
and an angle $\beta_2 = 60^{\circ}$ in the $z$-direction. 
A grid was generated as shown below:
\begin{equation}
   y_{i,j,k} = \Delta y (j - 1), \quad 
   x_{i,j,k} = \Delta x (i - 1) + \frac{y_{i,j,k}}{\tan(\beta_1)}, \quad 
   z_{i,j,k} = \Delta z (i - 1) + \frac{y_{i,j,k}}{\tan(\beta_2)},
\end{equation}
where $\Delta x$, $\Delta y$, and $\Delta z$ are the grid widths 
in the $x$-, $y$-, and $z$-directions, respectively. 
As the reference values used in this calculation, 
$l_\mathrm{ref} = L$ and $u_\mathrm{ref} = U$. 
The time step is changed from $\Delta t/(L/U) = 0.0005$ to 0.1, 
and we investigate the influence of the computational method on the time accuracy. 
The Courant number is defined as $\mathrm{CFL} = \Delta t U/\Delta x$ 
and changes as $\mathrm{CFL} = 0.005-1.0$ according to the time step. 
In this calculation, as an interpolation method of cell interface velocity, 
we use not only Rhie--Chow's interpolation but also direct interpolation, 
which does not use weighted interpolation by pressure.

\begin{figure}[!t]
\begin{minipage}[t]{0.49\hsize}
\begin{center}
\includegraphics[trim=0mm 0mm 0mm 0mm, clip, width=70mm]{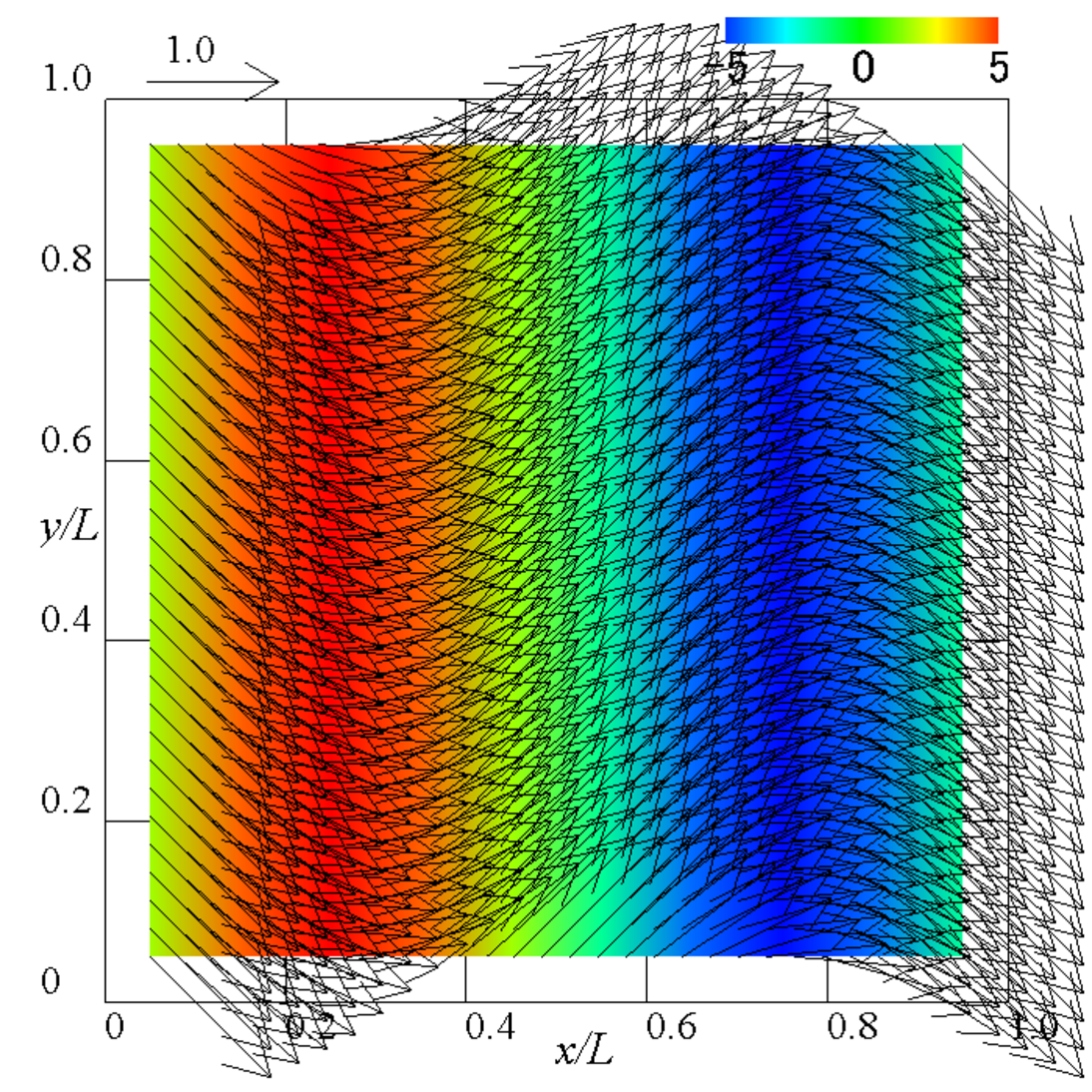} \\
(a) $t/(L/U) = 0$
\end{center}
\end{minipage}
\begin{minipage}[t]{0.49\hsize}
\begin{center}
\includegraphics[trim=0mm 0mm 0mm 0mm, clip, width=70mm]{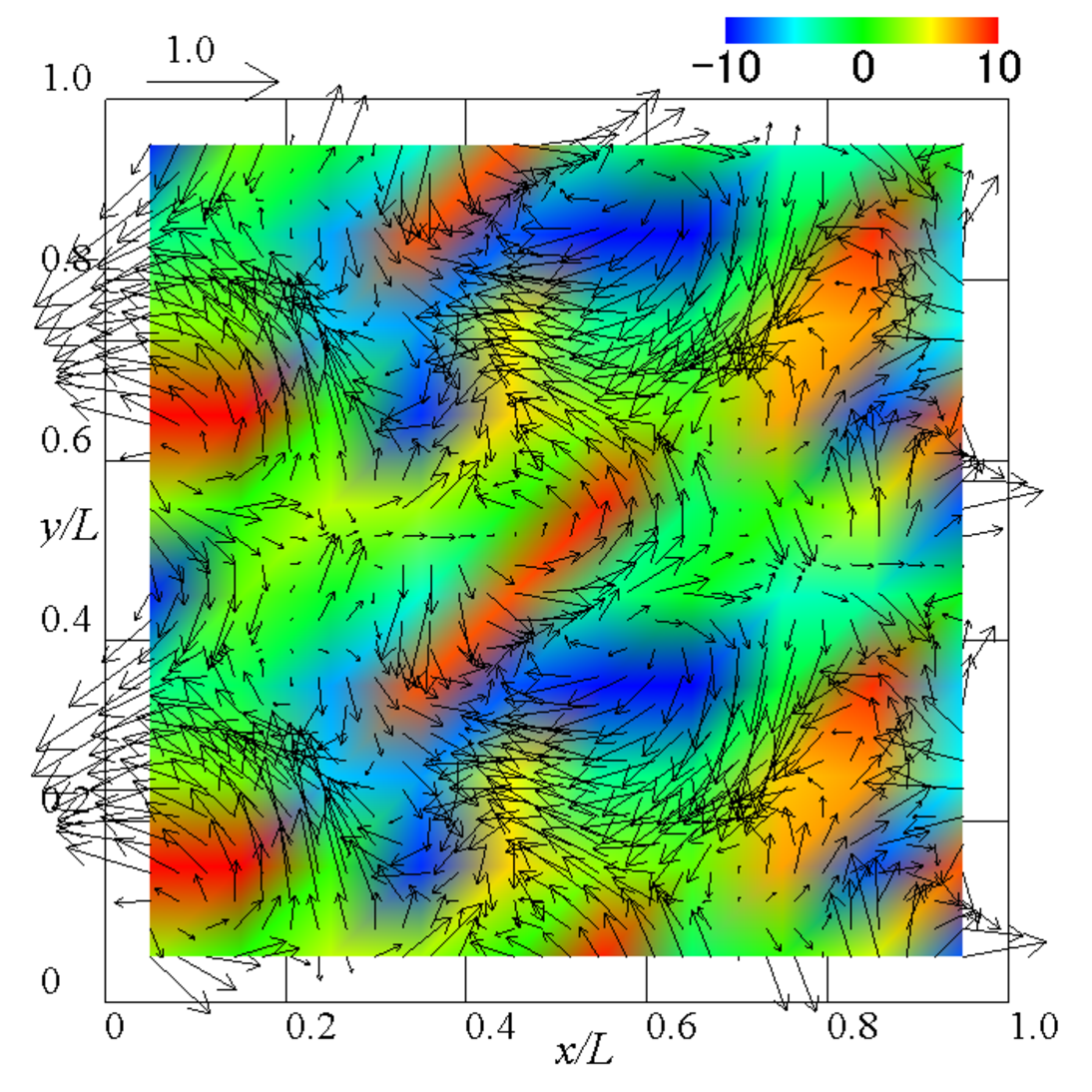} \\
(b) $t/(L/U) = 10$
\end{center}
\end{minipage}
\caption{Velocity vectors and vortisity contours at $z/L = 0.5$ 
for a periodic inviscid flow: $\Delta t/(L/U) = 0.01$}
\label{inviscid_v_wz}
\end{figure}

Figure \ref{inviscid_v_wz} shows velocity vectors and $z$-direction vorticity contours 
in the $x$-$y$ cross-section at times $t/(L/U) = 0$ and 10. 
The result at $t/(L/U) = 10$ was obtained using this numerical method. 
Although the initial state is a periodic flow field with wavelength $\lambda/L = 1.0$, 
at $t/(L/U) = 10$, the turbulence of twice the wave number component occurs, 
and the flow field develops into an unsteady and turbulent flow. 
It is found that, unlike the Taylor damping vortex, 
the initial state is not maintained.

\begin{figure}[!t]
\begin{minipage}[t]{0.49\hsize}
\begin{center}
\includegraphics[trim=0mm 0mm 0mm 0mm, clip, width=70mm]{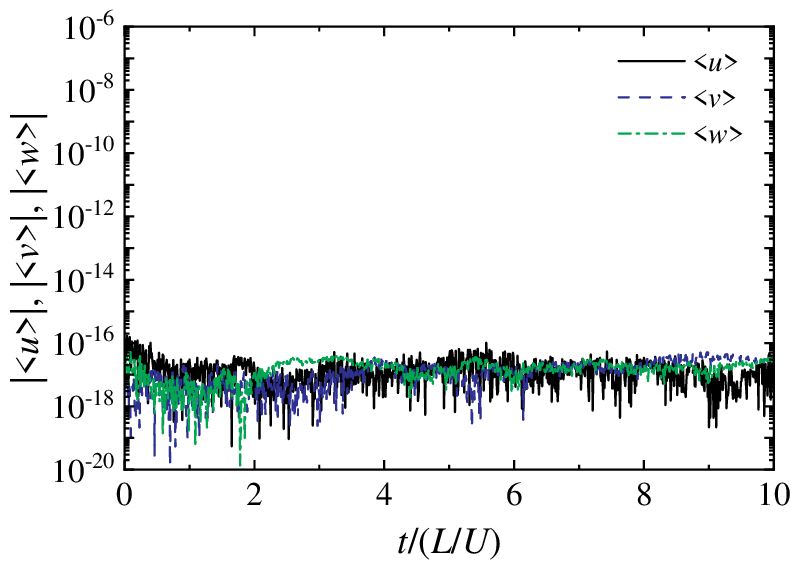} \\
(a) Velocity
\end{center}
\end{minipage}
\begin{minipage}[t]{0.49\hsize}
\begin{center}
\includegraphics[trim=0mm 0mm 0mm 0mm, clip, width=70mm]{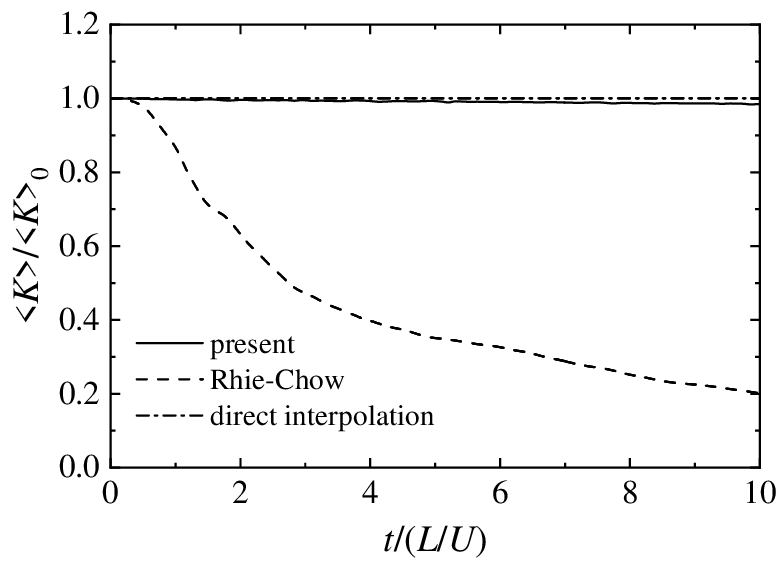} \\
(b) Kinetic energy
\end{center}
\end{minipage}
\caption{Total amounts of velocity and kinetic energy using orthogonal grid 
for a periodic inviscid flow: $\Delta t/(L/U) = 0.01$}
\label{inviscid_og_sum_u&K}
\end{figure}

As this computational model is a periodic flow, 
the total amount of velocity is conserved for $Re = \infty$. 
First, we consider the results using the orthogonal grid. 
Figure \ref{inviscid_og_sum_u&K}(a) shows the total amount, 
$\langle \bm{u} \rangle$, of velocity obtained using this numerical method. 
The total amounts for all velocity components change at low levels, 
indicating the excellent conservation of velocity. 
The total amount, $\langle K \rangle$, of kinetic energy is shown in Fig. \ref{inviscid_og_sum_u&K}(b). 
This calculated result almost agrees with that by direct interpolation, 
indicating that the energy is preserved. 
In the case of Rhie--Chow's interpolation, 
we can see that $\langle K \rangle$ decays rapidly, 
and the energy is not conserved.

\begin{figure}[!t]
\begin{center}
\includegraphics[trim=0mm 0mm 0mm 0mm, clip, width=70mm]{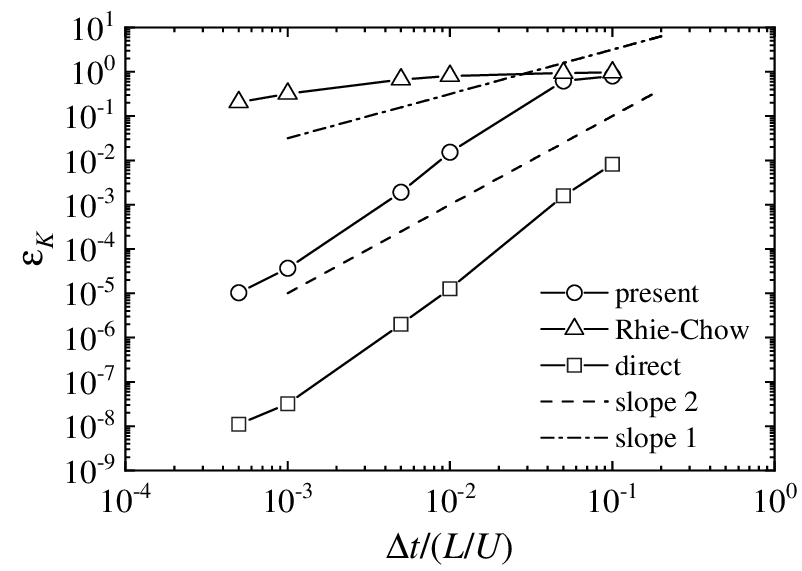}
\end{center}
\vspace*{-1.0\baselineskip}
\caption{Kinetic energy error using orthogonal grid 
for a periodic inviscid flow}
\label{inviscid_og_error}
\end{figure}

Figure \ref{inviscid_og_error} shows the relative error, 
$\varepsilon_{K} = |(\langle K \rangle-\langle K \rangle_0)/\langle K \rangle_0|$, 
of kinetic energy. 
Here, the subscript $0$ represents the initial value. 
As the time step $\Delta t$ decreases, 
the error $\varepsilon_{K}$ in this numerical method decreases with a slope of 2, 
similar to the direct interpolation method, 
indicating second-order convergence. 
In the direct interpolation without weighted interpolation by pressure, 
the error is lower than in the present scheme, 
and the weighted interpolation by pressure difference in this numerical method increases the error. 
The Rhie--Chow interpolation gives the kinetic energy a first-order accuracy error to time. 
However, Rhie--Chow interpolation does not show linear convergence, 
and lowering $\Delta t$ hardly reduces the error.

\begin{figure}[!t]
\begin{minipage}[t]{0.49\hsize}
\begin{center}
\includegraphics[trim=0mm 0mm 0mm 0mm, clip, width=70mm]{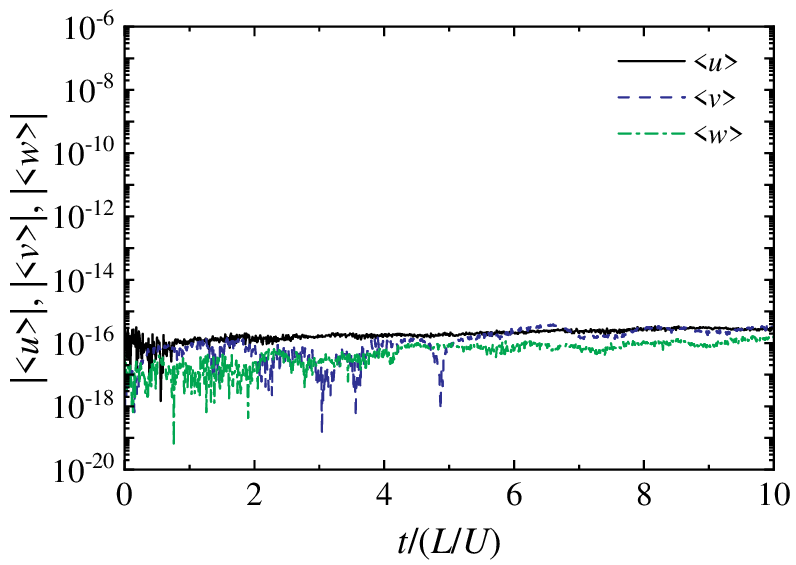} \\
(a) Velocity
\end{center}
\end{minipage}
\begin{minipage}[t]{0.49\hsize}
\begin{center}
\includegraphics[trim=0mm 0mm 0mm 0mm, clip, width=70mm]{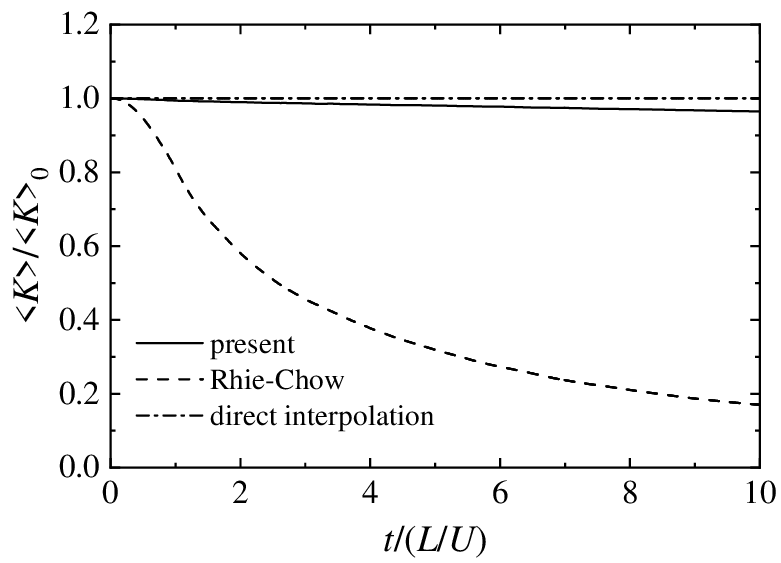} \\
(b) Kinetic energy
\end{center}
\end{minipage}
\caption{Total amounts of velocity and kinetic energy using skewed grid 
for a periodic inviscid flow: $\Delta t/(L/U) = 0.01$}
\label{inviscid_sg_sum_u&K}
\end{figure}

Next, we consider the results using the non-orthogonal grid. 
Figure \ref{inviscid_sg_sum_u&K}(a) shows the total amount, 
$\langle \bm{u} \rangle$, of velocity obtained using this computational method. 
As in the case of the orthogonal grid, the total amount is at the rounding error level, 
and the conservation of velocity is excellent. 
The total amount, $\langle K \rangle$, of kinetic energy is shown in Fig. \ref{inviscid_sg_sum_u&K}(b). 
Compared with the results by direct interpolation, 
the total amount obtained by this calculation method decreases slightly with time, 
and the energy conservation deteriorates. 
The Rhie--Chow interpolation does not conserve the kinetic energy.

\begin{figure}[!t]
\begin{center}
\includegraphics[trim=0mm 0mm 0mm 0mm, clip, width=70mm]{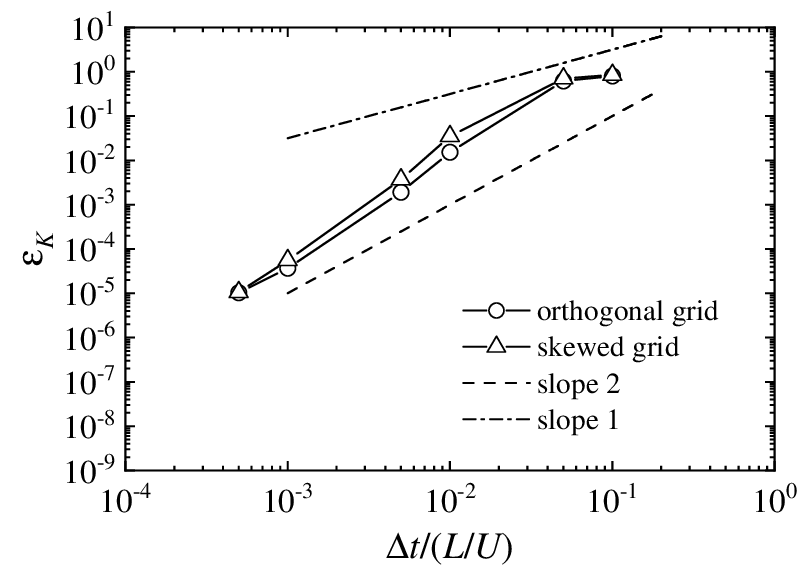}
\end{center}
\vspace*{-1.0\baselineskip}
\caption{Kinetic energy errors using orthogonal and skewd grids 
for a periodic inviscid flow}
\label{inviscid_og_sg_error}
\end{figure}

Figure \ref{inviscid_og_sg_error} shows the relative error, 
$\varepsilon_{K} = |(\langle K \rangle-\langle K \rangle_0)/\langle K \rangle_0|$, 
of kinetic energy. 
Even with the non-orthogonal grid, the error is at the same level as with the orthogonal grid. 
No deterioration in calculation accuracy is observed, 
and the second-order accuracy is maintained.

In this analysis, we investigated the inviscid flow. 
Even when analyzing unsteady flows with high Reynolds numbers, 
the Rhie--Chow interpolation method is expected to cause a significant attenuation 
of kinetic energy. 
It is believed that the improved scheme constructed in this study is effective 
in suppressing the attenuation of kinetic energy.

\section{Conclusion}
\label{summary}

This study constructed a finite difference scheme for incompressible fluids 
using a collocated grid in a general curvilinear coordinate system. 
The velocity at the cell interface is determined by weighted interpolation 
based on the pressure difference to prevent pressure oscillations. 
The Poisson equation for the pressure correction value is solved 
with the cross-derivative term omitted to improve calculation efficiency. 
In addition, simultaneous relaxation of velocity and pressure is applied 
to improve convergence. 
We analyzed steady flow fields to verify the validity of this numerical method. 
Even without the cross-derivative term, calculations could be stably performed, 
and convergent solutions were obtained. 
We investigated the conservation property of kinetic energy in unsteady flow fields 
and verified the accuracy of the cell interface interpolation method developed in this study. 
In inviscid flow, this computational method can suppress the attenuation of kinetic energy. 
In addition, it was revealed that the conservation of kinetic energy is excellent 
even with a non-orthogonal grid and 
that the calculation results have second-order accuracy to time, regardless of the grid. 
In the viscous analysis, when the Reynolds number increased, 
the error in this numerical method was lower than that of the Rhie--Chow interpolation method. 
It has been clarified that the present numerical scheme improves calculation accuracy in unsteady flows. 
In the future, we will apply this computational method to numerical analyses of 
coherent vortices and heat transport at high Reynolds number flows.

\section*{Acknowledgment}

The numerical results in this research were obtained 
using supercomputing resources at the Cyberscience Center, Tohoku University. 
This research did not receive any specific grant from funding agencies 
in the public, commercial, or not-for-profit sectors. 
We would like to express our gratitude to Associate Professor Yosuke Suenaga 
of Iwate University for his support of our laboratory. 
The authors wish to acknowledge the time and effort of everyone involved in this study.

\section*{Author declarations}

\noindent
{\bf Conflicts of interest}: The authors have no conflicts to disclose.

\noindent
{\bf Author contributions}: 
Hideki Yanaoka: Conceptualization (lead); Data curation (lead); 
Formal analysis (lead); Investigation (lead); Methodology (lead); 
Software (lead); Validation (lead); Visualization (lead); 
Writing – original draft (lead); Writing – review and editing (lead).

\bibliographystyle{arXiv_elsarticle-harv}
\bibliography{arXiv2023_yanaoka_bibfile}

\begin{thebibliography}{27}
\expandafter\ifx\csname natexlab\endcsname\relax\def\natexlab#1{#1}\fi
\providecommand{\url}[1]{\texttt{#1}}
\providecommand{\href}[2]{#2}
\providecommand{\path}[1]{#1}
\providecommand{\DOIprefix}{doi:}
\providecommand{\ArXivprefix}{arXiv:}
\providecommand{\URLprefix}{URL: }
\providecommand{\Pubmedprefix}{pmid:}
\providecommand{\doi}[1]{\href{http://dx.doi.org/#1}{\path{#1}}}
\providecommand{\Pubmed}[1]{\href{pmid:#1}{\path{#1}}}
\providecommand{\bibinfo}[2]{#2}
\ifx\xfnm\relax \def\xfnm[#1]{\unskip,\space#1}\fi
\bibitem[{Amsden and Harlow(1970)}]{Amsden&Harlow_1970}
\bibinfo{author}{Amsden, A.A.}, \bibinfo{author}{Harlow, F.H.},
  \bibinfo{year}{1970}.
\newblock \bibinfo{title}{A simplified {MAC} technique for incompressible fluid
  flow calculations}.
\newblock \bibinfo{journal}{J. Comput. Phys.} \bibinfo{volume}{6},
  \bibinfo{pages}{322--325}.
\newblock \DOIprefix\doi{https://doi.org/10.1016/0021-9991(70)90029-X}.
\bibitem[{Barakos et~al.(1994)Barakos, Mitsoulis and
  Assimacopoulos}]{Barakos_et_al_1994}
\bibinfo{author}{Barakos, G.}, \bibinfo{author}{Mitsoulis, E.},
  \bibinfo{author}{Assimacopoulos, D.}, \bibinfo{year}{1994}.
\newblock \bibinfo{title}{Natural convection flow in a square cavity revisited:
  {L}aminar and turbulent models with wall functions}.
\newblock \bibinfo{journal}{Int. J. Numer. Methods Fluids}
  \bibinfo{volume}{18}, \bibinfo{pages}{695--719}.
\newblock \DOIprefix\doi{https://doi.org/10.1002/fld.1650180705}.
\bibitem[{Bartholomewa et~al.(2018)Bartholomewa, Dennera, Abdol-Azisa, Marquisa
  and G.M.{van Wachema}}]{Bartholomew_et_al_2018}
\bibinfo{author}{Bartholomewa, P.}, \bibinfo{author}{Dennera, F.},
  \bibinfo{author}{Abdol-Azisa, M.H.}, \bibinfo{author}{Marquisa, A.},
  \bibinfo{author}{G.M.{van Wachema}, B.}, \bibinfo{year}{2018}.
\newblock \bibinfo{title}{Unified formulation of the momentum-weighted
  interpolation for collocated variable arrangements}.
\newblock \bibinfo{journal}{J. Comput. Phys.} \bibinfo{volume}{375},
  \bibinfo{pages}{177--208}.
\newblock \DOIprefix\doi{https://doi.org/10.1016/j.jcp.2018.08.030}.
\bibitem[{Choi(1999)}]{Choi_1999}
\bibinfo{author}{Choi, S.K.}, \bibinfo{year}{1999}.
\newblock \bibinfo{title}{Note on the use of momentum interpolation method for
  unsteady flows}.
\newblock \bibinfo{journal}{Numer. Heat Tr. A-Appl.} \bibinfo{volume}{36},
  \bibinfo{pages}{545--550}.
\newblock \DOIprefix\doi{https://doi.org/10.1080/104077899274679}.
\bibitem[{{De Vahl Davis}(1983)}]{Davis_1983}
\bibinfo{author}{{De Vahl Davis}, G.}, \bibinfo{year}{1983}.
\newblock \bibinfo{title}{Natural convection of air in a square cavity: {A}
  bench mark numerical solution}.
\newblock \bibinfo{journal}{Int. J. Numer. Methods Fluids} \bibinfo{volume}{3},
  \bibinfo{pages}{249--264}.
\newblock \DOIprefix\doi{https://doi.org/10.1002/fld.1650030305}.
\bibitem[{Fuchs and Tillmark(1985)}]{Fuchs&Tillmark_1985}
\bibinfo{author}{Fuchs, L.}, \bibinfo{author}{Tillmark, N.},
  \bibinfo{year}{1985}.
\newblock \bibinfo{title}{Numerical and experimental study of driven flow in
  polar cavity}.
\newblock \bibinfo{journal}{Int. J. Numer. Mthods Fluids} \bibinfo{volume}{5},
  \bibinfo{pages}{311--329}.
\newblock \DOIprefix\doi{https://doi.org/10.1002/fld.1650050403}.
\bibitem[{Harlow and Welch(1965)}]{Harlow&Welch_1965}
\bibinfo{author}{Harlow, F.H.}, \bibinfo{author}{Welch, J.E.},
  \bibinfo{year}{1965}.
\newblock \bibinfo{title}{Numerical calculation of time-dependent viscous
  incompressible flow of fluid with free surface}.
\newblock \bibinfo{journal}{Phys. Fluids} \bibinfo{volume}{8},
  \bibinfo{pages}{2182--2189}.
\newblock \DOIprefix\doi{https://doi.org/10.1063/1.1761178}.
\bibitem[{Hirt et~al.(1975)Hirt, Nichols and Romero}]{Hirt_et_al_1975}
\bibinfo{author}{Hirt, C.W.}, \bibinfo{author}{Nichols, B.D.},
  \bibinfo{author}{Romero, N.C.}, \bibinfo{year}{1975}.
\newblock \bibinfo{title}{{SOLA}: {A} numerical solution algorithm for
  transient fluid flows}.
\newblock \bibinfo{type}{Technical Report} \bibinfo{number}{LA-5852}. Los
  Alamos Scientific Lab., N. Mex.(USA).
\newblock \DOIprefix\doi{https://doi.org/10.2172/4205348}.
\bibitem[{Kim and Moin(1985)}]{Kim&Moin_1985}
\bibinfo{author}{Kim, J.}, \bibinfo{author}{Moin, P.}, \bibinfo{year}{1985}.
\newblock \bibinfo{title}{Application of a fractional-step method to
  incompressible {N}avier--{S}tokes equations}.
\newblock \bibinfo{journal}{J. Comput. Phys.} \bibinfo{volume}{59},
  \bibinfo{pages}{308--323}.
\newblock \DOIprefix\doi{https://doi.org/10.1016/0021-9991(85)90148-2}.
\bibitem[{Lee et~al.(2019)Lee, Jung, Kang and Hur}]{Lee_et_al_2019}
\bibinfo{author}{Lee, W.}, \bibinfo{author}{Jung, E.}, \bibinfo{author}{Kang,
  S.}, \bibinfo{author}{Hur, N.}, \bibinfo{year}{2019}.
\newblock \bibinfo{title}{On a momentum interpolation scheme for collocated
  meshes with improved discrete kinetic energy conservation}.
\newblock \bibinfo{journal}{J. Mech. Sci. Technol.} \bibinfo{volume}{33},
  \bibinfo{pages}{2761--2768}.
\newblock \DOIprefix\doi{https://doi.org/10.1007/s12206-019-0522-8}.
\bibitem[{Majumdar(1988)}]{Majumdar_1988}
\bibinfo{author}{Majumdar, S.}, \bibinfo{year}{1988}.
\newblock \bibinfo{title}{Role of underrelaxation in momentum interpolation for
  calculation of flow with nonstaggered grids}.
\newblock \bibinfo{journal}{Numer. Heat Tr.} \bibinfo{volume}{13},
  \bibinfo{pages}{125--132}.
\newblock \DOIprefix\doi{https://doi.org/10.1080/10407788808913607}.
\bibitem[{Morinishi(1998)}]{Morinishi_1998}
\bibinfo{author}{Morinishi, Y.}, \bibinfo{year}{1998}.
\newblock \bibinfo{title}{Fully conservative higher order finite difference
  schemes for incompressible flow}.
\newblock \bibinfo{journal}{J. Comput. Phys.} \bibinfo{volume}{143},
  \bibinfo{pages}{90--124}.
\newblock \DOIprefix\doi{https://doi.org/10.1006/jcph.1998.5962}.
\bibitem[{Morinishi(1999)}]{Morinishi_1999}
\bibinfo{author}{Morinishi, Y.}, \bibinfo{year}{1999}.
\newblock \bibinfo{title}{Improvement of collocated finite difference scheme
  with regard to kinetic energy conservation}.
\newblock \bibinfo{journal}{JSME, Ser. B} \bibinfo{volume}{65},
  \bibinfo{pages}{505--512}.
\newblock \DOIprefix\doi{https://doi.org/10.1299/kikaib.65.505}.
  \bibinfo{note}{(in Japanese)}.
\bibitem[{Patankar and Spalding(1972)}]{Patankar&Spalding_1972}
\bibinfo{author}{Patankar, S.V.}, \bibinfo{author}{Spalding, D.B.},
  \bibinfo{year}{1972}.
\newblock \bibinfo{title}{A calculation procedure for heat, mass and momentum
  transfer in three-dimensional parabolic flows}.
\newblock \bibinfo{journal}{Int. J. Heat Mass Transf.} \bibinfo{volume}{15},
  \bibinfo{pages}{1787--1806}.
\newblock \DOIprefix\doi{https://doi.org/10.1016/0017-9310(72)90054-3}.
\bibitem[{Peri{\'{c}}(1990)}]{Peric_1990}
\bibinfo{author}{Peri{\'{c}}, M.}, \bibinfo{year}{1990}.
\newblock \bibinfo{title}{Analysis of pressure-velocity coupling on
  nonorthogonal grids}.
\newblock \bibinfo{journal}{Numer. Heat Tr. B-Fund.} \bibinfo{volume}{17},
  \bibinfo{pages}{63--82}.
\newblock \DOIprefix\doi{https://doi.org/10.1080/10407799008961733}.
\bibitem[{Peri{\'{c}} et~al.(1988)Peri{\'{c}}, Kessler and
  Scheuerer}]{Peric_et_al_1988}
\bibinfo{author}{Peri{\'{c}}, M.}, \bibinfo{author}{Kessler, R.},
  \bibinfo{author}{Scheuerer, G.}, \bibinfo{year}{1988}.
\newblock \bibinfo{title}{Comparison of finite-volume numerical methods with
  staggered and colocated grids}.
\newblock \bibinfo{journal}{Comput. Fluids} \bibinfo{volume}{16},
  \bibinfo{pages}{389--403}.
\newblock \DOIprefix\doi{https://doi.org/10.1016/0045-7930(88)90024-2}.
\bibitem[{Rhie and Chow(1983)}]{Rhie&Chow_1983}
\bibinfo{author}{Rhie, C.M.}, \bibinfo{author}{Chow, W.L.},
  \bibinfo{year}{1983}.
\newblock \bibinfo{title}{Numerical study of the turbulent flow past an airfoil
  with trailing edge separation}.
\newblock \bibinfo{journal}{AIAA J.} \bibinfo{volume}{21},
  \bibinfo{pages}{1525--1532}.
\newblock \DOIprefix\doi{https://doi.org/10.2514/3.8284}.
\bibitem[{Rosenfeld et~al.(1991)Rosenfeld, Kwak and
  Vinokur}]{Rosenfeld_et_al_1991}
\bibinfo{author}{Rosenfeld, M.}, \bibinfo{author}{Kwak, D.},
  \bibinfo{author}{Vinokur, M.}, \bibinfo{year}{1991}.
\newblock \bibinfo{title}{A fractional step solution method for the unsteady
  incompressible {N}avier--{S}tokes equations in generalized coordinate
  systems}.
\newblock \bibinfo{journal}{J. Comput. Phys.} \bibinfo{volume}{94},
  \bibinfo{pages}{102--137}.
\newblock \DOIprefix\doi{https://doi.org/10.1016/0021-9991(91)90139-C}.
\bibitem[{Takemitsu(1985)}]{Takemitsu_1985}
\bibinfo{author}{Takemitsu, N.}, \bibinfo{year}{1985}.
\newblock \bibinfo{title}{Finite difference method to solve incompressible
  fluid flow}.
\newblock \bibinfo{journal}{J. Comput. Phys.} \bibinfo{volume}{61},
  \bibinfo{pages}{499--518}.
\newblock \DOIprefix\doi{https://doi.org/10.1016/0021-9991(85)90077-4}.
\bibitem[{Taylor(1923)}]{Taylor_1923}
\bibinfo{author}{Taylor, G.I.}, \bibinfo{year}{1923}.
\newblock \bibinfo{title}{Lxxv. on the decay of vortices in a viscous fluid}.
\newblock \bibinfo{journal}{The London, Edinburgh, and Dublin Philosophical
  Magazine and Journal of Science, Series 6} \bibinfo{volume}{46},
  \bibinfo{pages}{671--674}.
\newblock \DOIprefix\doi{https://doi.org/10.1080/14786442308634295}.
\bibitem[{{Van der Vorst}(1992)}]{Vorst_1992}
\bibinfo{author}{{Van der Vorst}, H.A.}, \bibinfo{year}{1992}.
\newblock \bibinfo{title}{{Bi--CGSTAB}: {A} fast and smoothly converging
  variant of {Bi--CG} for the solution of nonsymmetric linear systems}.
\newblock \bibinfo{journal}{SIAM J. Sci. and Stat. Comput.}
  \bibinfo{volume}{13}, \bibinfo{pages}{631--644}.
\newblock \DOIprefix\doi{https://doi.org/10.1137/0913035}.
\bibitem[{{Van Doormaal} and Raithby(1984)}]{Van_Doormaal&Raithby_1984}
\bibinfo{author}{{Van Doormaal}, J.P.}, \bibinfo{author}{Raithby, G.D.},
  \bibinfo{year}{1984}.
\newblock \bibinfo{title}{Enhancements of the {SIMPLE} method for predicting
  incompressible fluid flows}.
\newblock \bibinfo{journal}{Numer. Heat Tr.} \bibinfo{volume}{7},
  \bibinfo{pages}{147--163}.
\newblock \DOIprefix\doi{https://doi.org/10.1080/01495728408961817}.
\bibitem[{Wu et~al.(1995)Wu, Squires and Wang}]{Wu_et_al_1995}
\bibinfo{author}{Wu, X.}, \bibinfo{author}{Squires, K.D.},
  \bibinfo{author}{Wang, Q.}, \bibinfo{year}{1995}.
\newblock \bibinfo{title}{Extension of the fractional step method to general
  curvilinear coordinate systems}.
\newblock \bibinfo{journal}{Numer. Heat Tr. B- Fund.} \bibinfo{volume}{27},
  \bibinfo{pages}{175--194}.
\newblock \DOIprefix\doi{https://doi.org/10.1080/10407799508914952}.
\bibitem[{Yanaoka(2023)}]{Yanaoka_2023}
\bibinfo{author}{Yanaoka, H.}, \bibinfo{year}{2023}.
\newblock \bibinfo{title}{Influences of conservative and non-conservative
  {L}orentz forces on energy conservation properties for incompressible
  magnetohydrodynamic flows}.
\newblock \bibinfo{journal}{J. Comput. Phys.} \bibinfo{volume}{491},
  \bibinfo{pages}{112372 (36 pages)}.
\newblock \DOIprefix\doi{https://doi.org/10.1016/j.jcp.2023.112372}.
\bibitem[{Yanaoka and Inafune(2023)}]{Yanaoka&Inafune_2023}
\bibinfo{author}{Yanaoka, H.}, \bibinfo{author}{Inafune, R.},
  \bibinfo{year}{2023}.
\newblock \bibinfo{title}{Frequency response of three-dimensional natural
  convection of nanofluids under microgravity environments with gravity
  modulation}.
\newblock \bibinfo{journal}{Numer. Heat Tr. A-Appl.} \bibinfo{volume}{83},
  \bibinfo{pages}{745--769}.
\newblock \DOIprefix\doi{https://doi.org/10.1080/10407782.2022.2161437}.
\bibitem[{Yu et~al.(2002)Yu, Kawaguchi, Tao and Ozoe}]{Yu_et_al_2002}
\bibinfo{author}{Yu, B.}, \bibinfo{author}{Kawaguchi, Y.},
  \bibinfo{author}{Tao, W.Q.}, \bibinfo{author}{Ozoe, H.},
  \bibinfo{year}{2002}.
\newblock \bibinfo{title}{Checkerboard pressure predictions due to the
  underrelaxation factor and time step size for a nonstaggered grid with
  momentum interpolation method}.
\newblock \bibinfo{journal}{Numer. Heat Tr. B- Fund.} \bibinfo{volume}{41},
  \bibinfo{pages}{85--94}.
\newblock \DOIprefix\doi{https://doi.org/10.1080/104077902753385027}.
\bibitem[{Zang et~al.(1994)Zang, Street and Koseff}]{Zang_et_al_1994}
\bibinfo{author}{Zang, Y.}, \bibinfo{author}{Street, R.L.},
  \bibinfo{author}{Koseff, J.R.}, \bibinfo{year}{1994}.
\newblock \bibinfo{title}{A non-staggered grid, fractional step method for
  time-dependent in compressible {N}avier--{S}tokes equations in curvilinear
  coordinates}.
\newblock \bibinfo{journal}{J. Comput. Phys.} \bibinfo{volume}{114},
  \bibinfo{pages}{18--33}.
\newblock \DOIprefix\doi{https://doi.org/10.1006/jcph.1994.1146}.

\end{thebibliography}


\end{document}